\documentclass[%
journal=jacsat,manuscript=article,
amsmath,amssymb,amsthm,mathrsfs
]{achemso}
\usepackage[version=3]{mhchem} 
\usepackage{bm}  
\usepackage{url}
\usepackage{hyperref}
\usepackage{enumitem}
\usepackage[normalem]{ulem}
\usepackage[usenames,dvipsnames]{xcolor}
\setlist[enumerate]{label*=\arabic*.}

\author{Benjamin A. Frandsen}
\email{benfrandsen@byu.edu}
\affiliation[Brigham Young University]
{Department of Physics and Astronomy, Brigham Young University, Provo, Utah 84602, United States of America}
\author{Henry E. Fischer}
\affiliation[Institut Laue-Langevin]{Institut Laue-Langevin, 71 avenue des Martyrs, CS 20156, 38042 Grenoble cedex 9, France}
\email{fischer@ill.fr}

\title[]
  {A New Spin on Material Properties: Local Magnetic Structure in Functional and Quantum Materials}

\abbreviations{IR,NMR,UV}

\begin{document}

\def\ie{{\em i.e.\/}}
\def\eg{{\em e.g.\/}}
\def\etal{{\em et al.\/}}
\def\deltampdf{3D-$\Delta$mPDF}
\def\TN{$T_{\mathrm{N}}$}
\def\TC{$T_{\mathrm{C}}$}

%
\def\simlt{\mathrel{\mathpalette\oversim<}}
\def\simgt{\mathrel{\mathpalette\oversim>}}
\hyphenation{
diffracto-gramme diffracto-grammes
regard-less
}
%

%
%
%
%
%

\begin{abstract}
The past few decades have made clear that the properties and performances of emerging functional and quantum materials can depend strongly on their local atomic and/or magnetic structure, particularly when details of the local structure deviate from the long-range structure averaged over space and time. Traditional methods of structural refinement (\eg~Rietveld) are typically sensitive only to the average structure, creating a need for more advanced structural probes suitable for extracting information about structural correlations on short length- and time-scales. In this Perspective, we describe the importance of local magnetic structure in several classes of emerging materials and present the magnetic pair distribution function (mPDF) technique as a powerful tool for studying short-range magnetism from neutron total-scattering data. We then provide a selection of examples of mPDF analysis applied to magnetic materials of recent technological and fundamental interest, including the antiferromagnetic semiconductor MnTe, geometrically frustrated magnets, and iron-oxide magnetic nanoparticles. The rapid development of mPDF analysis since its formalization a decade ago puts this technique in a strong position for making continued impact in the study of local magnetism in emerging materials.

\end{abstract}


\section{Introduction}

\subsection{Why worry about local structure?}


``Structure determines function'' is a familiar adage in the study of materials, and for good reason---for decades, this principle has guided and motivated structural studies in contexts ranging from proteins to metal-organic frameworks to topological insulators. In materials chemistry and physics, this phrase is typically understood to refer to the crystallographic, long-range-ordered structure as averaged over space and time.  Indeed, global symmetries corresponding to the crystallographic structure provide a powerful framework for understanding and predicting material properties. Together with the fact that material property calculations based on density functional theory usually require the assumption of a perfectly periodic structure, the emphasis on the crystallographic structure is understandable.

On the other hand, global symmetries and crystallographic structure do not always tell the whole story. For a large and growing list of materials of technological and fundamental importance, the \textit{local} atomic structure is a key determinant of macroscopic properties~\cite{billi;cc04, dagot;s05, young;jmc11, keen;n15, zhu;advs21, zunge;ncs22, kimbe;nm23}. By local atomic structure, we mean correlations between atoms that are well defined on short spatial and temporal scales but may average away to zero on longer scales. Therefore, the local instantaneous structure may be quite different from the time-averaged long-range (\ie~space-averaged) structure, e.g. with symmetries that are broken on the nanoscale but preserved globally. The features of the local structure can often profoundly influence material properties. Examples include the unusual negative thermal expansion in \ce{ZrW2O8}\cite{Tucker+etal2005}, efficiency and performance in Li-ion battery materials~\cite{wang;stam17}, the size of the band gap in semiconductors~\cite{liu;acsn22}, and the exceptionally large electromechanical response in relaxor ferroelectrics~\cite{li;afm18}, to name a few. On the basis of these examples and many others, the importance of gaining a detailed understanding of the local atomic structure of materials is now widely accepted.



\subsection{Why worry about local \textit{magnetic} structure?}

\begin{figure}
    \includegraphics[width=80mm]{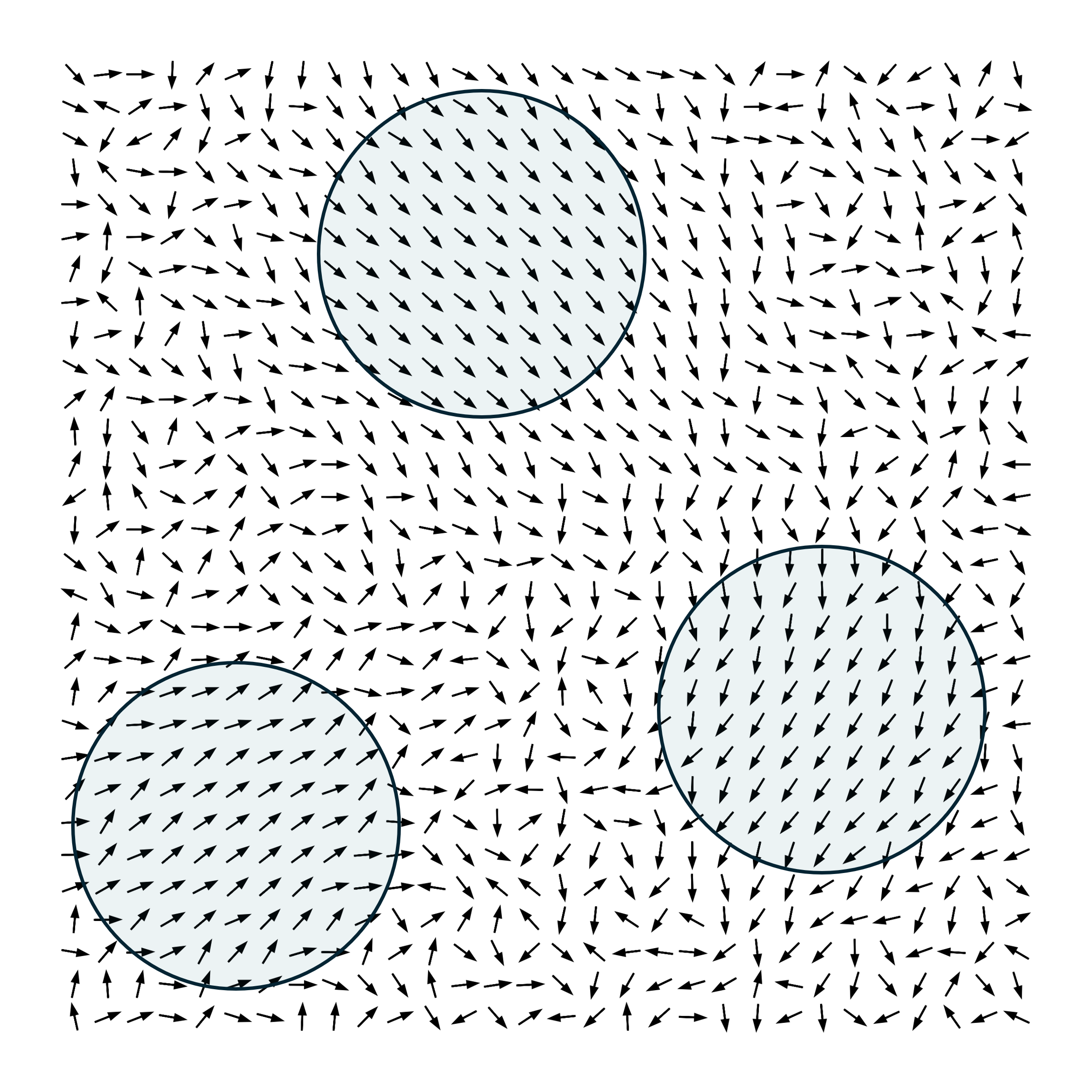}
    \centering
    \caption{Schematic illustration of short-range ferromagnetic correlations in a two-dimensional square lattice. The local ferromagnetic configuration is well-defined within ferromagnetic domains indicated by the circled regions having diameter comparable to the spin correlation length.  As the distance between two spins increases, the ferromagnetic correlation is progressively lost, ultimately decreasing to zero when the relative spin orientation of two distant spins becomes fully random.  Such a short spin correlation length as depicted here would not produce observable magnetic Bragg peaks in a neutron diffraction experiment,
    {\color{black}
    but only broad bumps of magnetic diffuse-scattering intensity.
    }
    }
    \label{fig:SRO}
\end{figure}

More recently, there has been increased recognition that the local magnetic structure, \ie~the arrangement of atomic magnetic moments (colloquially called spins) on short spatial and temporal scales, can also be critically important for understanding the behavior of magnetic materials. 
Short-range magnetic correlations such as those illustrated in Fig.~\ref{fig:SRO} can arise in a variety of situations and materials, either as static or dynamically fluctuating correlations. Examples include the ``correlated paramagnet'' regime observed in many magnetic materials at temperatures just above a magnetic ordering transition, where short-range, metastable magnetic structure persists~\cite{paddi;prb18, Qureshi+etal2022}; magnets with strong chemical disorder~\cite{fisch;prb02}; nonmagnetic host materials with dilute magnetic impurities~\cite{dietl;rmp14}; and materials such as spin glasses and spin liquids with intrinsically disordered magnetic ground states~\cite{mydos;rpp15, clark;armr21}. Even in well-behaved, conventional magnets, whenever the ordered magnetic moment in Rietveld analysis is found to be smaller than the full local magnetic moment of a magnetic ion, the ``disordered component'' of the magnetic structure can in fact manifest significant local spin correlations that may be static or dynamic in nature.

We mention two reasons why this type of local magnetic structure is important. First, local magnetic correlations can directly impact macroscopic materials properties, such as the thermopower in magnetic semiconductors~\cite{zheng;sadv19}, the entropy change in magnetocalorics~\cite{miao;rm18}, the stabilization of high-entropy materials~\cite{walsh;pnas21}, and even the performance of catalysts for water electrolysis for hydrogen production~\cite{minne;aprev24}. Second, the local magnetic structure often contains vital clues about the underlying physics of the material, such as magnetoelectric coupling in multiferroics~\cite{jones;prb24}, detailed information about interactions between spins~\cite{Qureshi+etal2022}, the nature of phase transitions~\cite{bean;pr62}, or even the mechanism of unconventional superconductivity~\cite{norma;s11}. Establishing a comprehensive understanding of not just the average magnetic structure, but also the local magnetic structure, is therefore imperative in the study of many magnetic materials of technological and fundamental relevance.

In this Perspective, we explore these ideas further by offering a selection of recent examples showing the importance of local magnetic structure, then describing how local magnetic structure can be probed experimentally through magnetic pair distribution function (mPDF) analysis of neutron total-scattering data~\cite{Frandsen+etal2014}, and finally presenting a few illustrative examples of mPDF analysis in topical materials. Considering the growing scientific need for quantitative studies of local magnetic structure and the emerging experimental techniques well suited to meet this need, we see tremendous potential for growth and progress in this area in the near future.


\section{Local magnetic structure in emerging materials: A selection of recent examples}


Here, we briefly highlight a few materials that have garnered recent interest to illustrate how local magnetic structure can influence material properties and/or shed light on the underlying physics. We focus on quantum materials and energy-conversion materials, although many other types of materials could also be discussed in this context. The intent here is not to provide a comprehensive review of any one material or class of materials, but instead to illustrate the breadth of materials and applications for which local magnetic structure is relevant. 

\subsection{Quantum materials}

Materials with electronic and magnetic properties that originate from nontrivial quantum mechanical effects are collectively referred to as ``quantum materials''~\cite{cava;cr21}. Prominent examples of quantum materials include strongly correlated electron systems such as high-$T_c$ copper oxide superconductors~\cite{keime;n15}, geometrically frustrated magnets with exotic magnetic ground states~\cite{clark;armr21}, and electronic topological materials like topological insulators and Weyl semimetals~\cite{naran;nm21}.

\begin{figure}
    \centering
    \includegraphics[width=120mm
    ]{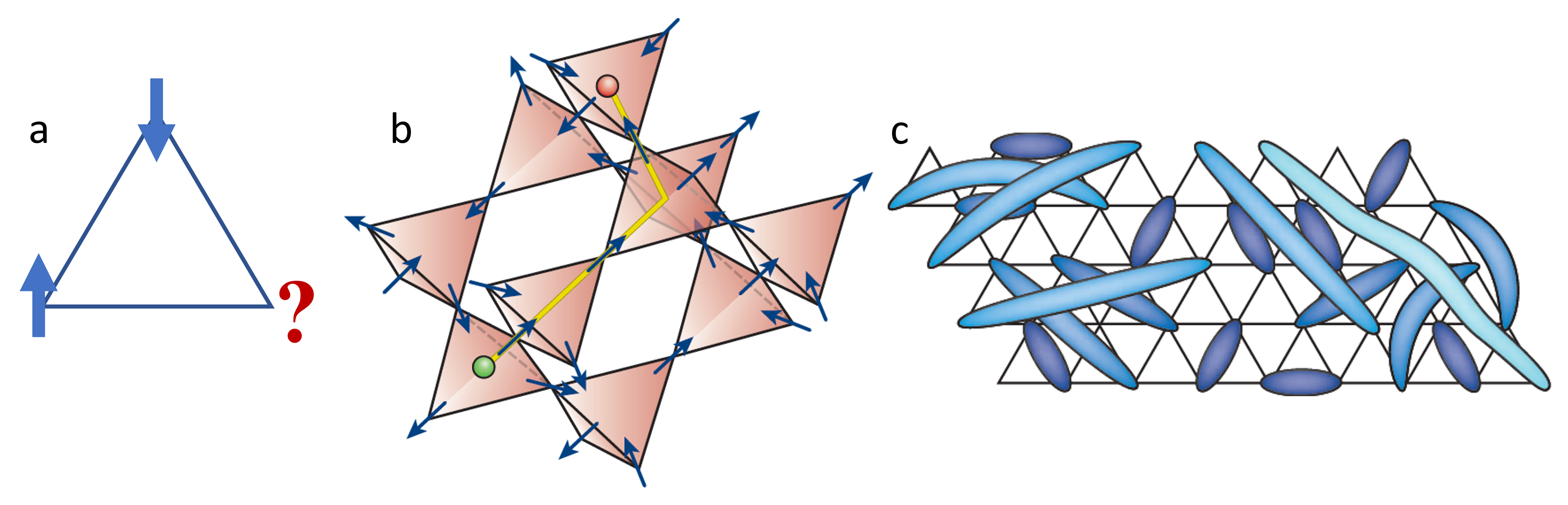}
    \caption{Geometrical frustration in quantum magnets. (a) Antiferromagnetically coupled spins on a triangular lattice form an archetypal motif for geometrical frustration. (b) Flipping one spin in the spin-ice ground state creates a pair of emergent magnetic monopoles (green and red spheres) that are weakly bound in the lattice by a magnetic Coulomb force. (c) Schematic representation of one possible configuration of maximally entangled spin pairs on a triangular lattice. A ``resonating valence bond'' quantum spin liquid (QSL) arises as a quantum superposition of all such configurations.  Panels (b, c) adapted with permission from Ref.~\cite{balen;n10}. Copyright 2010 Springer Nature.}
    \label{fig:frustration-basics}
\end{figure}

 Many illustrations of the importance of local magnetic structure can be found among quantum materials, with geometrically frustrated magnets providing perhaps the most obvious example. In these materials, the geometry of the crystal structure prevents competing magnetic interactions from being simultaneously satisfied~\cite{ramir;aroms94}. A quintessential example is found in antiferromagnetically coupled spins on a triangular lattice, for which it is impossible for all three spins on a given triangle to be mutually antiparallel (Fig.~\ref{fig:frustration-basics}a).
Frustration suppresses the tendency for long-range magnetic order to form, yet the magnetic interactions often favor some form of short-range order. A beautiful illustration  of this is seen in the spin ice compounds \ce{Ho2Ti2O7} and \ce{Dy2Ti2O7}~\cite{caste;arocmp12}, in which large rare-earth magnetic moments occupy a pyrochlore network of corner-sharing tetrahedra (Fig.~\ref{fig:frustration-basics}b). Strong Ising anisotropy forces the moments to point directly toward or away from the center of the tetrahedron, which frustrates the dipolar interactions between the moments. No long-range magnetic order is established even at zero temperature; instead, the system freezes into one of a macroscopically large number of degenerate configurations in which each tetrahedron has exactly two moments pointing inward and two pointing outward, in precise analogy to the arrangement of protons in water ice. The two-in-two-out ``ice rule'' therefore establishes a highly correlated local magnetic structure even in the absence of long-range magnetic order. Remarkably, the low-energy excitations of a spin ice are analogous to a gas of weakly interacting emergent magnetic monopoles (Fig.~\ref{fig:frustration-basics}b)~\cite{caste;arocmp12}. This rich new physics hinges on the details of the local magnetic structure in spin ice compounds. 

The most exotic ground state for a geometrically frustrated magnet is a quantum spin liquid (QSL), in which the effects of frustration and quantum fluctuations are so strong that the system remains dynamically fluctuating even at zero temperature~\cite{balen;n10, savar;rpp17}. A QSL exhibits many-body entanglement of spins in the absence of conventional long-range order, although short-range spatial and temporal correlations are present~\cite{broho;s20}. One conceptual model of a QSL proposed by Philip Anderson, known as a resonating valence bond QSL~\cite{ander;mrb73}, is shown schematically in Fig.~\ref{fig:frustration-basics}c. QSLs and QSL-adjacent states have attracted intense interest as ideal platforms for fundamental studies of many-body quantum entanglement and even potential applications for quantum computation~\cite{knoll;arocmp19}. Detailed knowledge of the local magnetic correlations in such a state provides crucial information about the nature of the ground state and the level of quantum entanglement~\cite{musto;cm24, Nuttall+etal2023, schei;np24}.




Magnetic van der Waals (vdW) materials, which have recently emerged as an exciting new class of quantum materials~\cite{burch;n18,wang;adp20,yang;advs21,wang;acsn21,ren;nanom23}, provide another example of local magnetic structure influencing material properties in crucial ways. Van der Waals materials typically have a layered crystal structure with weak vdW interactions between the layers, lending the materials a strongly two-dimensional (2D) electronic character dominated by in-plane interactions. They exist as three-dimensional bulk crystals and can also be exfoliated or grown as mono- or few-layer crystals. Van der Waals systems with intrinsic magnetism provide an ideal playground for exploring the intersection of low dimensionality, magnetism, and topology, and have attracted intense interest for both fundamental science and applications in next-generation spintronic, magnetoelectric, and magneto-optic devices. According to the Mermin-Wagner theorem~\cite{mermi;prl66}, long-range magnetic order for an isotropic Heisenberg magnetic system is destroyed by thermal fluctuations for $d\le2$, where $d$ is the spatial dimension, although short-range correlations are expected. Magnetic anisotropy (e.g. Ising, XY, or magnetocrystalline anisotropy) in vdW magnets introduces gaps into the magnetic excitation spectrum, thus stabilizing magnetic order against thermal fluctuations and permitting long-range order. Nevertheless, thermal fluctuations still play a crucial role in the physics of vdW magnets, and short-range magnetic order is widely observed in the otherwise paramagnetic state of vdW systems, indicating that local magnetic structure is important in these materials.

\begin{figure}
    \centering
    \includegraphics[width=150mm]{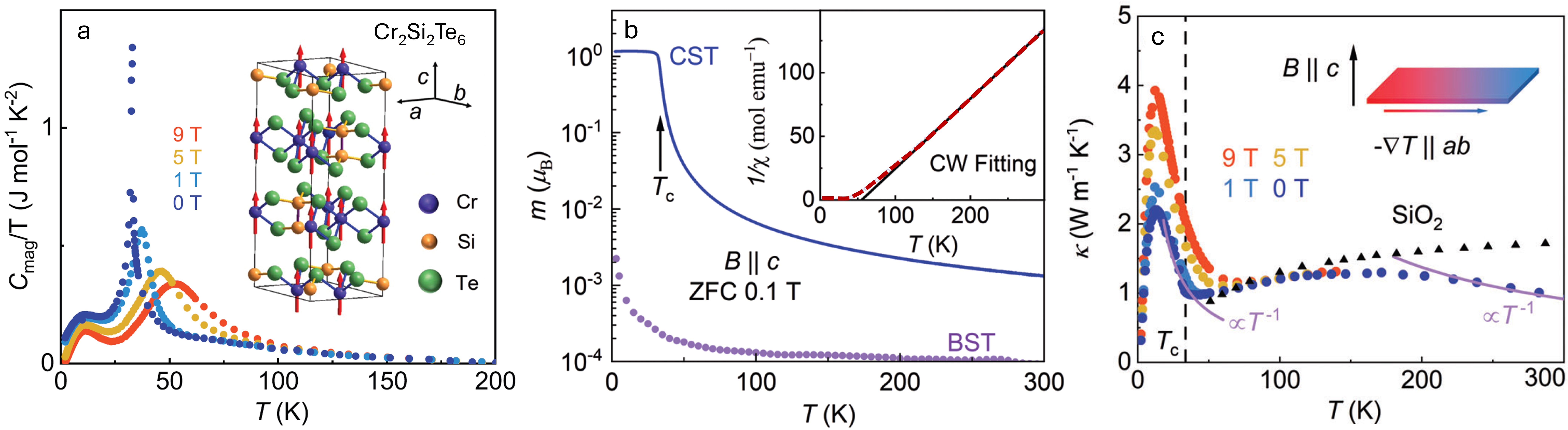}
    \caption{ Magnetic heat capacity (a), magnetization (b), and thermal conductivity (c) of \ce{Cr2Si2Te6} as a function of temperature and applied magnetic field. The insets to (a) and (b) show the crystal structure and temperature-dependent inverse magnetic susceptibility, respectively. The magnetic heat capacity was isolated from the total heat capacity by subtracting the calculated phonon contribution determined from first-principles calculations. Adapted with permission from Ref.~\citenum{yang;afm23}. Copyright 2023 John Wiley and Sons.}
    \label{fig:Cr2Si2Te6}
\end{figure}

A specific illustration of this is found in the thermal conductivity of the insulating vdW ferromagnet \ce{Cr2Si2Te6} (CST)~\cite{yang;afm23}, the crystal structure of which is shown in the inset of Fig.~\ref{fig:Cr2Si2Te6}(a). The Cr$^{3+}$ spins exhibit long-range order below 33~K, as seen by the sharp peak in the heat capacity and rapid increase in magnetization shown in Fig.~\ref{fig:Cr2Si2Te6}(a,b).
 In addition, pronounced short-ranged correlations survive up to about 200~K~\cite{willi;prb15}, as evidenced by the nonzero magnetic heat capacity (panel a) and deviation from linear Curie-Weiss behavior (inset of panel b) below that temperature scale.  These correlated magnetic fluctuations that persist far above the ordering temperature are enhanced by the 2D nature of the magnetism. The thermal conductivity $\kappa$ highlights the profound impact of this short-range magnetism. As shown in Fig.~\ref{fig:Cr2Si2Te6}(c), $\kappa$ stays around approximately 1~W/m/K above \TC\ and reaches a peak of only about 2~W/m/K below \TC\ in zero field. This is a remarkably low thermal conductivity for a crystalline material, comparable to amorphous silica (black triangles). Moreover, the temperature dependence changes from $1/T$ above 200~K, which is the expected behavior\cite{KeyesPR1959} for normal phonon transport dominated by Umklapp scattering, to a decreasing trend upon further cooling from 200~K to \TC, precisely in the temperature region where short-range magnetism is most pronounced. This can be explained by spin-phonon scattering, whereby phonons scatter from regions of fluctuating, short-range magnetic correlations. As the temperature decreases, the correlation length and number density of the magnetic fluctuations grow, thus scattering phonons more effectively and decreasing the thermal conductivity. This scenario is confirmed by the increase in $\kappa$ with applied magnetic field, which partially polarizes the spins and therefore suppresses the short-range spin fluctuations, leading to reduced spin-phonon scattering and higher thermal conductivity. In this way, a key macroscopic property that is a crucial consideration for many potential devices based on vdW magnets depends strongly on the local magnetic structure and can be tuned by manipulating the short-range spin correlations.


\subsection{Energy-conversion materials}


Magnetism plays an important role in several classes of energy-conversion materials. One example is found in magnetocaloric materials, for which the application or removal of a magnetic field causes a temperature change resulting from the exchange of entropy between the lattice and spin subsystems. For example, when a ferromagnetically ordered system is demagnetized and enters a disordered magnetic state, the magnetic entropy increases. If this occurs adiabatically, then the increase in magnetic entropy must be compensated by a decrease in the entropy of another part of the system, usually the lattice. With a decrease in lattice entropy, the temperature of the material correspondingly decreases. The potential of exploiting this magnetocaloric effect (MCE) in a cyclic fashion has ignited intense recent interest in the context of thermal energy harvesting and alternative solid-state refrigeration technologies, which have the potential to be more efficient than conventional vapor-based refrigeration and obviate the need for refrigerant gases with high global warming potential~\cite{kitan;aem20}. 
Magnetic cooling~\cite{Magnetic_Cooling_in_LowT_Book_1988} at low temperature is also highly useful for applications such as helium liquefaction.

Recently, so-called giant MCE materials have emerged as particularly promising candidates for applications~\cite{franc;armr12}. 
Under zero applied magnetic field as a function of temperature, these materials exhibit a first-order magnetostructural or magnetoelastic transition in which the discontinuous nature of the magnetic transition results in an exceptionally large magnetic entropy change in a narrow temperature window around the transition, leading to large magnetic cooling power. However, this comes at the cost of a concomitant first-order structural transition, typically with significant hysteresis and a large change in unit cell dimensions, which can lead to mechanical failure after repeated cycles. Developing the ability to minimize hysteresis and discontinuous structural changes while maintaining the large magnetic entropy change is therefore a top priority for giant MCE materials. In that context, (Mn,Fe)$_2$(P,Si,B) is an appealing system~\cite{miao;rm18}, because varying the stoichiometry allows one to tune the ferromagnetic transition in terms of both the transition temperature and the character of the transition (\ie~first-order to second-order). This tunability raises the possibility of designing an optimal giant MCE material. 

\begin{figure}
    \centering
    \includegraphics[width=120mm]{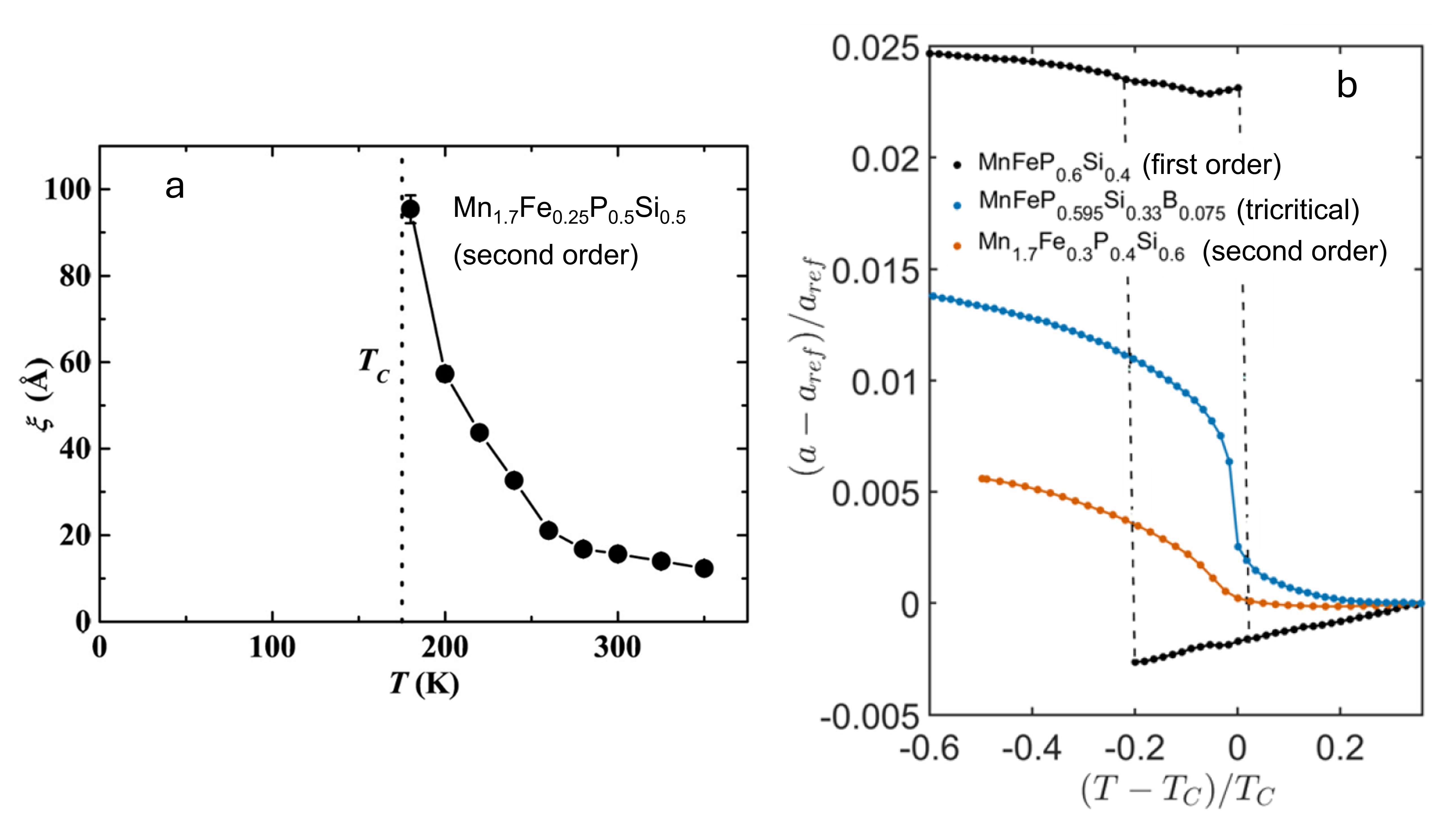}
    \caption{ (a) Short-range magnetic correlation length in magnetocaloric compound Mn$_{1.7}$Fe$_{0.25}$P$_{0.5}$Si$_{0.5}$ extracted from muon spin relaxation and neutron scattering data. Adapted with permission from~\cite{miao;prb16}. Copyright 2016 American Physical Society. (b) Magnetostructural lattice parameter change in various compositions of (Mn,Fe)$_2$(P,Si,B). Adapted from~\cite{boeij;jpd17}. Available under a CC-BY license. Copyright 2017 M.F.J. Boeije \etal}
    \label{fig:MnFePSi}
\end{figure}

Importantly, short-range magnetic correlations are thought to be intimately tied to the tunability of (Mn,Fe)$_2$(P,Si,B). In compositions with second-order transitions, robust short-range magnetic correlations survive up to  high temperatures deep into the nominally paramagnetic state~\cite{miao;prb16}, as seen in Fig.~\ref{fig:MnFePSi}(a).
In contrast, short-range magnetism is greatly suppressed or entirely absent above the magnetostructural transition in the first-order compositions. Additionally, the lattice parameter change in first-order compositions is discontinuous, hysteretic, and much larger than that observed in second-order compositions or compositions near the tricritical point~\cite{boeij;jpd17}, as seen in Fig.~\ref{fig:MnFePSi}(b). The gentler temperature evolution of the lattice in the tricritical and second-order compositions is attributed to the influence of the short-range magnetic correlations in the nominally paramagnetic state~\cite{boeij;jpd17}.
In this way, local magnetic correlations help control crucial material properties in MCE compounds. In some cases, short-range magnetism alone can drive a large MCE, even in the complete absence of long-range magnetic order~\cite{pakhi;sr17, munir;ic20}.

Thermoelectric materials, in which a thermal gradient results in a spontaneously generated voltage, constitute another important class of energy-conversion materials with tremendous potential for application. 
Thermoelectric performance is typically quantified by the dimensionless figure of merit $zT=\sigma S^2 T/\kappa$, where $\sigma$ is the electrical conductivity, $S$ is the Seebeck coefficient, $T$ is temperature, and $\kappa$ is the thermal conductivity~\cite{shi;cr20}. Materials with $zT$ of 1 or above are typically considered high-performance thermoelectrics that could be viable for practical applications. As seen in the formula for $zT$, good thermoelectrics should have a large Seebeck coefficient and electrical conductivity but low thermal conductivity. Unfortunately, this combination of material properties is rare due to the natural correlation between $S$, $\sigma$, and $\kappa$, requiring creative efforts to optimize $zT$.

One of the most significant developments in thermoelectric research over the last few years has been the realization that the spin degree of freedom offers numerous routes toward higher $zT$ that had been largely overlooked during the previous decades of research~\cite{yang;nrp23, liu;advpr23}. This greatly broadens the scope of candidate high-performance thermoelectric materials and has led to novel device design concepts, such as transverse thermoelectrics based on the Nernst effect~\cite{kumar;mcf23}. For many mechanisms of magnetically enhanced thermoelectricity, local magnetic structure plays a key role. For example, the Seebeck coefficient quantifies the average entropy flow per charge carrier as induced by a temperature gradient, such that the entropy flow associated with disordered spin orientations and/or degenerate spin-orbital configurations in magnetic conductors can often increase $S$ without significantly reducing $\sigma$ or increasing $\kappa$, thus increasing $zT$~\cite{koshi;prl01}. Additionally, spin fluctuations near a magnetic phase transition or short-range correlations caused by frustration often strongly impact transport and thermodynamic properties in ways that can significantly increase $zT$~\cite{polas;crps21, wang;prb23}, \eg~by reducing $\kappa$ as shown in Fig.~\ref{fig:Cr2Si2Te6}(c). Later in this Perspective, we will discuss in detail the example of MnTe~\cite{zheng;sadv19, baral;matter22}, in which thermally induced fluctuations of short-range magnetic correlations dramatically increase $zT$ at high temperature. In short, utilizing local magnetic structure to enhance thermoelectric properties is among the most promising avenues for developing next-generation thermoelectric materials.




\section{How do we probe local magnetic structure?}

\begin{figure}
    \centering
    \includegraphics[width=120mm]{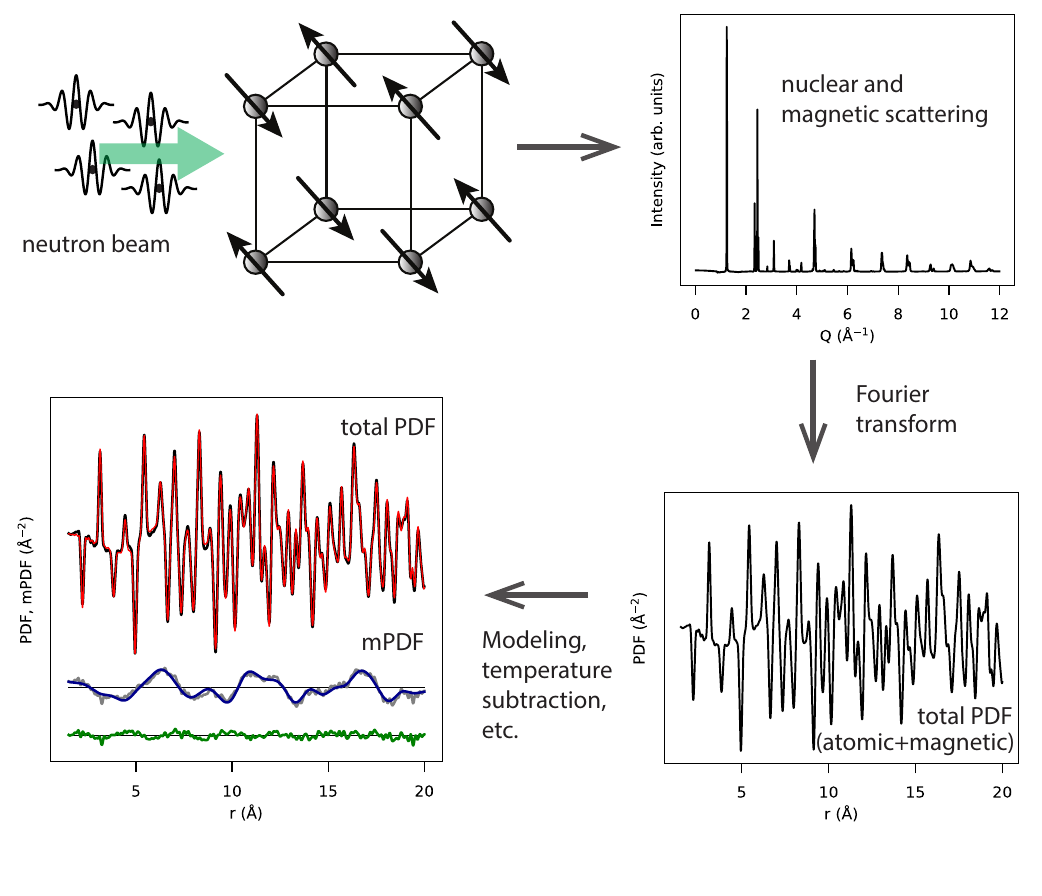}
    \caption{ Schematic summary of magnetic PDF analysis. Adapted with permission from Ref.~\citenum{billi;chapter;cic23}. Copyright 2023 Elsevier.}
    \label{fig:schematic}
\end{figure}

Having established the importance of local magnetic structure, we now address the challenge of studying it experimentally. In general, any description of a material's structure must specify the length and time scales under consideration and any spatial/temporal averages being made in that description.  Likewise, any structural measurement, such as x-ray or neutron diffraction, probes specific length and time scales and includes various averaging effects. Therefore, the spatial/temporal scales of modelled and measured structures must match to enable a proper comparison. 
In this Perspective, we present mPDF analysis as a powerful new tool for probing magnetic structure on local length- and local time-scales that are relevant for understanding the properties of many types of magnetic materials. A schematic overview of the mPDF approach is shown in Fig.~\ref{fig:schematic}, with details provided later in this section.
In essence, the method entails performing a neutron diffraction experiment and Fourier transforming the total-scattering diffraction intensity to produce the real-space atomic/magnetic pair distribution function (PDF), from which detailed information about the local atomic and magnetic correlations can be extracted through direct inspection of the data and/or detailed modeling. In the remainder of this section, we briefly introduce atomic PDF analysis to provide context for mPDF analysis, outlining the formalism of both, and finally address some practical aspects of mPDF experimentation and data analysis.

\subsection{Basics of atomic PDF analysis}

Atomic PDF analysis is a method of extracting local structure information from neutron or x-ray scattering experiments that goes beyond the average structure information available through conventional Bragg diffraction analysis. Here, we briefly introduce the technique for the case of neutron scattering, while noting that several thorough reviews of the formalism and applications of atomic PDF analysis can be found in the literature.\cite{Egami+Billinge_book_2012, Fischer+etal_Review_2006, Terban+Billinge_2022, Keen2020, Peterson+Keen_2021}
In a scattering experiment conducted on an isotropic sample (\eg~glass, liquid, powdered crystal) with a collimated beam of monochromatic neutrons with incident wavelength $\lambda$, the diffracted intensity (counts/s) measured by a detector cell at scattering angle $2\theta$ can be written as
\begin{equation}
 I(q)
 =\Phi\,\,\frac{d\sigma}{d\Omega}(q)\,\,d\Omega,
\end{equation}
where the scalar $q =|\bm{q}|=(4\pi/\lambda)\,\sin(\theta)$ is the wave-vector transfer, $\Phi$ is the incident flux, and $d\Omega$ is the solid angle of the detector cell.  The diffraction pattern is therefore proportional to the {\it differential scattering cross-section\/}~$d\sigma/d\Omega$, which is a function of $q$. Once normalized per atom, this can be written as the following sum of a $q$-dependent ``distinct'' term (interference between different atoms) and an isotropic ``self'' term (self-interference from individual atoms):
\begin{equation}
 \frac{1}{N}\,\left[\frac{d\sigma}{d\Omega}(q)\right]
 =\ \;\overline{b}^2\; [ S(q)-1 ] \ +\ \overline{b^2},
    \label{eqn:dsdo_self}
\end{equation}
where the dimensionless function $S(q)$ is called the {\it static structure factor\/} for the material and contains all the structural information of the diffraction pattern.  Here, $\overline{b}$ is the ``coherent'' scattering length averaged over all isotopes and nuclear spin states of the atomic species.


The local atomic structure of the sample is revealed via Fourier transform of the distinct term $S(q)-1$ over the experimentally accessible $q$-range [$q_{\rm min}, q_{\rm max}$], leading to the experimentally determined, model-independent PDF$(r)$,
\begin{equation}
 {\rm PDF}(r) 
 \ \ =\ \ \frac{2}{\pi}\int_{q_{\rm min}}^{q_{\rm max}}
\negthinspace\negthinspace\negthinspace
 q\; [S(q)-1]\; {\rm sin}(qr)\; dq
  \ \ =\ \ \frac{1}{N}\, 
 \sum_{i\not=j}^N
 \bigg[ \, \frac{1}{r} \delta(r-r_{ij})\bigg]
\, - \, 4\pi r\rho_0
 \ .
   \label{eqn:aPDF}
\end{equation}
The peaks arising from the $\delta$-functions occur at atomic pair separation distances $r_{ij}$, and integration of the peak areas leads to the average atomic coordination numbers; thus, the PDF provides a histogram of the local atomic structure. Our assumption of isotropic scattering makes the PDF an orientational (\ie~powder) average, where the first term is the standard expression for the spherically-averaged auto-correlation function for a volume of distinct point-like scatterers.
The second term is linear in $r$ with a negative slope proportional to the average atomic number density $\rho_0$.~\cite{Farrow+Billinge2009}
In a real experiment, the $\delta$-functions are broadened due to the finite value of $q_{\mathrm{max}}$, resulting in a finite real-space resolution~$\Delta r$. In addition, the real-space extent of the PDF signal is limited by the finite $q$-resolution $\Delta q$ of the diffraction experiment as determined by the coherence volume of the scattered neutrons.~\cite{Gaehler+etal1998} Larger $q_{\rm max}$ and finer $\Delta q$ lead to improved real-space resolution and increased real-space extent. Typical neutron PDF instruments provide $\Delta r$ in the range 0.1 -- 0.15 \AA\ and $r_{\rm max}$ in the range 50 -- 100 \AA.


An important note is that PDF$(r)$ is an {\it ensemble average\/} over all local atomic structures in the material. Moreover, for the vast majority of PDF experiments conducted on instruments that do not analyze the energy of the scattered beam (and thus effectively integrate over all energy transfers), the PDF probes atomic correlations with an extremely short neutron coherence time~\cite{Gaehler+etal1998} typically on the order of femtoseconds, much shorter than the fastest atomic vibration periods of 0.1~ps or so. Because the entire scattering pattern is considered (e.g. Bragg and diffuse scattering, elastic and inelastic scattering), atomic PDF analysis is referred to as a ``total scattering'' technique. Note that even a perfectly crystalline material will produce some diffuse and inelastic scattering due to thermal atomic vibrations.

In summary, a total-scattering diffraction pattern and its Fourier transform PDF($r$) 
{\it represent an ensemble average of quasi-instantaneous local structures within the sample\/}. By including all diffusely and inelastically scattered neutrons, it contains the maximal amount of structural information about the sample that is available via diffraction, including all dynamic and static positional correlations between neighboring atoms. 
In contrast, conventional analysis of integrated Bragg peak intensities (\eg~Rietveld refinement)  represents only the sample's space- and time-averaged structure, disregarding inter-peak diffuse intensity containing information about spatially local deviations from crystalline periodicity as well as inelastic scattering containing information about dynamic atomic correlations (\eg~phonons)\cite{Desgranges+etal_2023}. 

{\color{black}
We note that certain neutron spectrometers can function as diffractometers with a variable range of energy-exchange integration~$\Delta E$, thereby adjusting the coherence time over which the sample's structure is time-averaged by a couple orders of magnitude, which allows to measure the characteristic time-scales of the local structure's dynamics, as well as to identify the atomic displacements responsible for observed vibrational excitation energies.  This relatively new method is generally referred to ``dynamic PDF'' analysis\cite{kimbe;nm23, McQueeney1998, Fry-Petit2015}.
}

Although the preceding discussion has focused on neutron scattering, it applies largely to x-ray scattering as well, with the important caveat that x-ray scattering is subject to a $q$-dependent form factor that requires some additional consideration.

\subsection{Basics of magnetic PDF analysis}




Magnetic PDF analysis, which follows in close analogy to atomic PDF analysis, provides access to local magnetic correlations through the Fourier transform of the diffraction signal produced by neutrons scattering from magnetic moments in a material. We consider a system of $N$ identical atomic magnetic moments $\bm{\mu}_{\rm m}$, casually referred to as ``spins'', whose magnitude $\mu_{\rm m}$ is expressed in units of the Bohr magneton~$\mu_{\rm B}$ and represents the total local atomic magnetic moment including the spin-orbit interaction as well as itinerant electron contributions, \ie~the magnetic moment magnitude that contributes to magnetic scattering intensity.  Quantum mechanical interactions between the spins lead to orientational correlations, which may be quantified at a given instant via classical vector products rather than spin operators, following Wright\cite{Wedgwood+Wright1976,Wright1980} and others\cite{Keen+McGreevy1991,Paddison+etal_Spinvert_2013,Kodama+etal2017,Qureshi_Mag2Pol_2019}.  
This method allows us to simplify somewhat the expression first derived by Blech and Averbach\cite{Blech+Averbach1964} for the total {\it magnetic\/} differential unpolarized neutron scattering cross-section per atom of an isotropically-scattering system of $N$ spins (\eg~a powdered crystal),
and for zero magnetic field applied to the sample:
\begin{equation}
\frac{1}{N}\frac{d\sigma}{d\Omega}\bigg|_{\rm m} =\ \ p^2\mu_{\rm m}^2f_{\rm m}^2(q)\;
\Bigg\{ 
\frac{2}{3} + \frac{1}{N} \sum\limits_{i\not=j}
\bigg[ 
\hat{A}_{ij} \frac{\sin(qr_{ij})}{qr_{ij}} 
+ \hat{B}_{ij}\bigg(\frac{\sin(qr_{ij})}{(qr_{ij})^3}
- \frac{\cos(qr_{ij})}{(qr_{ij})^2} \bigg)
\bigg] 
\Bigg\} ,
  \label{eqn:BandA}
\end{equation}
where $p=\gamma_n r_e/2 = 2.696$~fm with $\gamma_n = 1.913$ as the neutron magnetic moment in units of nuclear magnetons and $r_e = 2.818$~fm as the classical electron radius, $f_{\rm m}(q)$ is the magnetic form factor with $f_{\rm m}(0)=1$, 
and $\hat{A}_{ij}$ and $\hat{B}_{ij}$ 
are spin orientational correlation coefficients for spin components that are respectively perpendicular (\ie~transverse) or predominantly parallel (\ie~longitudinal) to the interspin vector $\bm{r}_{ij} = \bm{r}_j - \bm{r}_i$.  
Defining unit vectors $\hat{\bm{\mu}}_{\rm m} = \bm{\mu}_{\rm m} / \mu_{\rm m}$ for the magnetic moments of spins $i$ and $j$, we can express their orientational correlation coefficients in terms of simple products of Cartesian components:
\begin{equation}
\label{eqn:AhatBhat}
\hat{A}_{ij} \ \buildrel{\rm def}\over{=}\ \hat{\mu}_{\rm m,i}^y \; \hat{\mu}_{\rm m,j}^y
\hskip+8em
\hat{B}_{ij} \ \buildrel{\rm def}\over{=}\ 2\;\hat{\mu}_{\rm m,i}^x \; \hat{\mu}_{\rm m,j}^x - \hat{A}_{ij}
\end{equation}
where our choice of local coordinate system aligns $\bm{x}=\bm{r}_{ij}/{r}_{ij}$ for each spin pair whose $(x,y)$ plane contains the $i$th unit spin vector $\hat{\mu}_{\rm m,i}$ (but not necessarily also $\hat{\mu}_{\rm m,j}$).


The first term on the right-hand side of Eq.~\ref{eqn:BandA} is the magnetic self-scattering per atom,
{\color{black}
\ie~the perfect orientational correlation of a given magnetic spin with itself,
}
which can be written as
\begin{equation}
\label{eqn:magscattlen}
\frac{d\sigma}{d\Omega}\bigg|_{\rm m,self} =\ b_{\rm m}^2(q)\; = \;\frac{2}{3}p^2\mu_{\rm m}^2f_{\rm m}^2(q),
\end{equation}
where we have introduced the magnetic scattering length $b_{\rm m}(q) \;\buildrel{\rm def}\over{=}\; \sqrt\frac{2}{3}\;p\;\mu_{\rm m}\;f_{\rm m}(q)$. The magnetic self-scattering corresponds to the total diffraction intensity per spin in the absence of orientational correlations between neighboring spins.  Such a sample with zero spin correlations is simply in a perfect paramagnetic state.
By subtracting the magnetic self-scattering $b^2_{\rm m}(q)$ from the total magnetic differential scattering cross-section, we retain only that corresponding to correlations between {\color{black} distinct magnetic spins ($i\not=j$),} \ie~the distinct term.  Also in analogy to atomic PDF analysis, this distinct term is then normalized through division by the magnitude of magnetic diffraction intensity per magnetic ion given by Eq.~\ref{eqn:magscattlen}. Typically, only the value at $q=0$ is used, \ie\ $\frac{2}{3}p^2\mu_{\rm m}^2 = b^2_{\mathrm{m}}(q=0)$, although in some cases dividing by the full $q$-dependent $b^2_{\mathrm{m}}(q)$ may be desirable (see further discussion below). For now, we divide by  $b^2_{\mathrm{m}}(q=0)$ and then take the Fourier transform to produce the {\it model-independent\/} mPDF($r$):\cite{Frandsen+etal2014}
 $$
{\rm mPDF}(r)
\ \buildrel{\rm def}\over{=}\ \frac{2}{\pi}\int_{q_{\rm min}}^{q_{\rm max}}
\hskip-0.5em q\; \frac{\frac{1}{N}\frac{d\sigma}{d\Omega}\big|_{\rm m} -\ b_{\rm m}^2(q)}{\frac{2}{3}p^2\mu_{\rm m}^2}\; {\rm sin}(qr)\; dq 
 $$
{\color{black}
\begin{equation}
\approx \frac{3}{2}   
\bigg\{
\frac{1}{N} 
\sum\limits_{i\not=j} 
\bigg[ \frac{\hat{A}_{ij}}{r} \widetilde{\delta}(r-r_{ij}) 
 + \hat{B}_{ij}\frac{r}{r_{ij}^3} \widetilde{\Theta}(r_{ij}-r) 
 \bigg]
 - 4\pi r\rho_{\rm m}\langle F \rangle / \mu_{\rm m}^2
 \bigg\}
 \ ,
 \label{eqn:mPDF_<F>}
\end{equation}
}
where $\rho_{\rm m}$ is the number density of magnetic ions and $\langle F \rangle = \frac{2}{3} \langle\bm{\mu}_{\rm m}\rangle\cdot\langle\bm{\mu}_{\rm m}\rangle$, \ie~the 
square of the average magnetic moment vector $\langle\bm{\mu}_{\rm m}\rangle$ within a magnetic domain~\cite{Wedgwood+Wright1976,Kodama+etal2017,Frandsen+etal_mpdf_2022} (which is zero for magnetic domains having no net magnetization, \eg\ antiferromagnets). {\color{black}
Substituting this expression for $\langle F \rangle$ allows us to simplify our expression somewhat to
\begin{equation}
{\rm mPDF}(r)
\;\approx\ \ \frac{3}{2}  
\frac{1}{N} 
\sum\limits_{i\not=j} 
\bigg[ \frac{\hat{A}_{ij}}{r} \widetilde{\delta}(r-r_{ij}) 
 + \hat{B}_{ij}\frac{r}{r_{ij}^3} \widetilde{\Theta}(r_{ij}-r)
\bigg]
\; -\; 4\pi r\rho_{\rm m} |\langle\hat{\bm{\mu}}_{\rm m}\rangle|^2
 \ ,
 \label{eqn:mPDF_mu2}
\end{equation}
}
which, as expected, is independent of the common magnetic moment magnitude $\mu_{\rm m}$ (again, expressed in units of Bohr magnetons $\mu_{\rm B}$). 
We note again that we have chosen to quantify the orientational correlations between spins via classical vector products.
Here we are also assuming that small-angle neutron scattering (SANS) from very long-range magnetic spin correlations is not included in the scattering intensity that is Fourier transformed.

As for the atomic PDF$(r)$ of Eq.~\ref{eqn:aPDF}, the first term in our result (Eq.~\ref{eqn:mPDF_mu2}) for the magnetic PDF$(r)$ is the spherically-averaged auto-correlation function of the spatial distribution of the scattering centers encountered by the unpolarized neutrons, appropriately weighted by the coefficient $\hat{A}_{ij}$.  Moreover, the term weighted by $\hat{B}_{ij}$ averages out to zero in the case of cubic symmetry\cite{Frandsen+etal2014} or when the relative orientation of spins~$i$ and~$j$ is independent of the {\it direction\/} of the interspin vector $\bm{r}_{ij}$.\cite{Naegele+etal1978,Wright1980} 
The mPDF$(r)$ even includes the subtraction of a term linear in $r$ (albeit zero for antiferromagnets), again similar to the atomic PDF$(r)$. However, in contrast to neutron scattering from point-like nuclei having a certain scattering length, which leads to the sum of $\delta$-functions in Eq.~\ref{eqn:aPDF} for the atomic PDF$(r)$, magnetic scattering intensity results from neutrons interacting with extended volumes of magnetic scattering-length density (\ie~spin density), giving rise to the magnetic form factor $f_{\rm m}(q)$.  The peaks in mPDF$(r)$ are therefore appropriately broadened by the spatial extent of the magnetic spin density surrounding the nuclei of the magnetic ions, as indicated by the tildes over the delta and Heaviside step functions above, and as observed experimentally when magnetic total-scattering intensity is Fourier transformed to mPDF$(r)$.  Approximating $f_{\rm m}^2(q)$ as a Gaussian (whence the $\approx$ signs in Eqs.~\ref{eqn:mPDF_<F>} and~\ref{eqn:mPDF_mu2}), this real-space broadening of the mPDF$(r)$ peaks is given by ${\rm FWHM}_{\rm R}\ \approx\ 4\ln(4)/{\rm FWHM}_{f_{\rm m}^2(q)}$.  In some cases however, it can be useful to go a step further and to divide the magnetic intensity additionally by $f_{\rm m}^2(q)$ before Fourier transformation, effectively performing a real-space deconvolution of the spin-density distribution, in order to sharpen the peaks of the mPDF$(r)$.\cite{Frandsen+Billinge2015} Note however that such a mathematical operation assumes isotropy in the spin density around magnetic ions and may require corrections for inaccuracies in the heuristically modelled magnetic form factor functions at high~$q$.

\begin{figure}
    \centering
    \includegraphics[width=100mm]{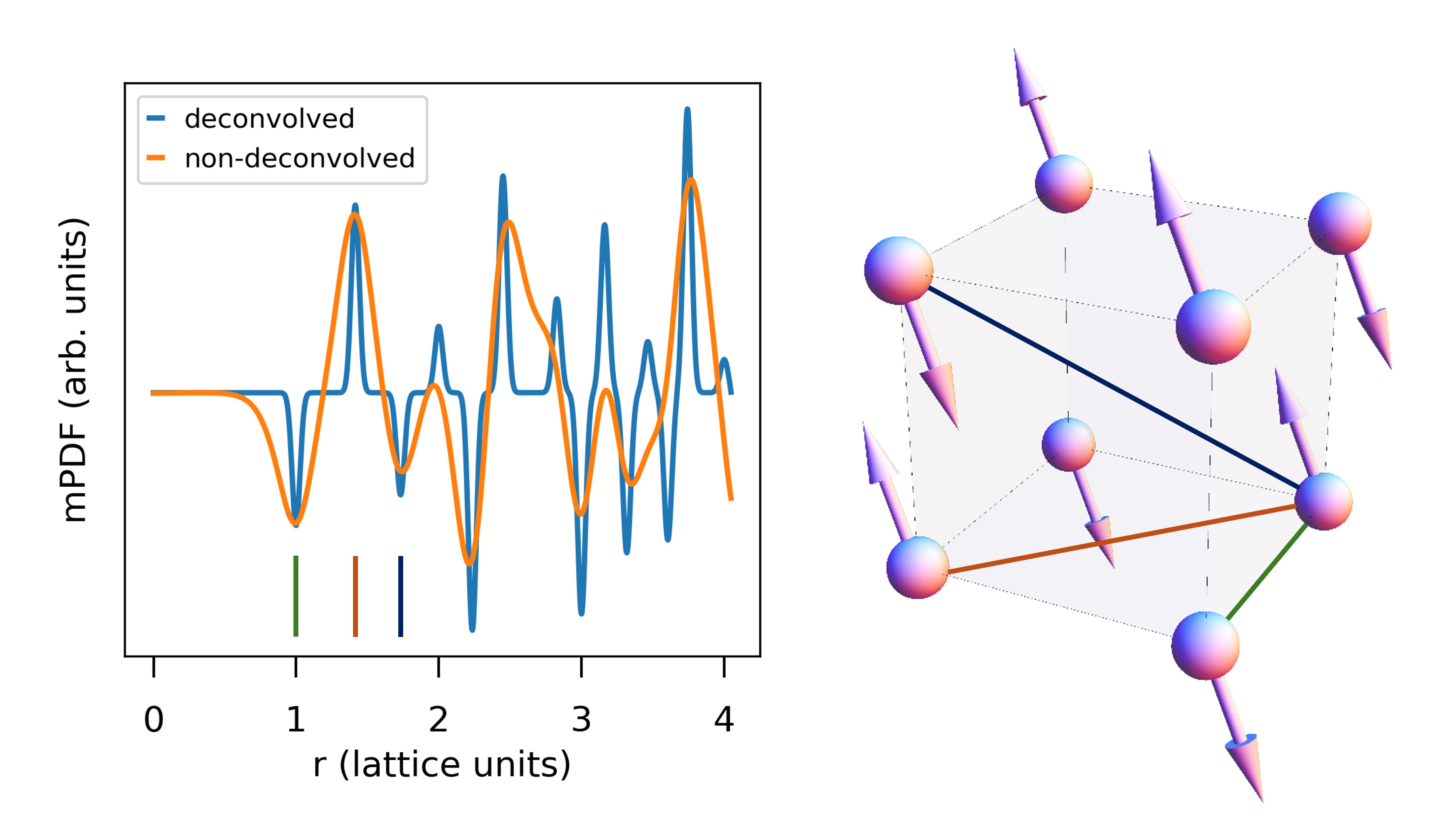}
    \caption{ \color{black}Simulated mPDF pattern (left) from a hypothetical simple cubic antiferromagnet (right) highlighting the intuitive nature of the mPDF. The ``deconvolved'' mPDF pattern represents the ideal result when the magnetic scattering intensity can be reliably divided by the squared magnetic form factor prior to performing the Fourier transform; this is not done for the ``non-deconvolved'' pattern, which reflects what is typically done in practice and is consistent with Eq.~\ref{eqn:mPDF_<F>}. Colored vertical lines on the left correspond to representative spin pairs shown by the same colors on the right.\color{black}
    }
    \label{fig:simplecubicAFM}
\end{figure}

Happily, Eq.~\ref{eqn:mPDF_mu2} is very intuitive, and shows that peaks arise in the mPDF pattern at spin-pair separation distances, and that the sign of the peak depends on whether the transverse spin correlations are parallel (positive) or antiparallel (negative). Direct inspection of the mPDF therefore provides a simple way to visualize the local magnetic structure, as illustrated for the hypothetical case of a simple cubic AFM lattice in Fig.~\ref{fig:simplecubicAFM}.
Further information about the longitudinal versus transverse correlations can be inferred from a more detailed look at the mPDF pattern, including the slope at low $r$, and/or by comparison with a simulated mPDF pattern from a given magnetic model. 

As discussed for atomic PDF, standard magnetic diffraction experiments effectively integrate over all possible energy exchanges between the sample and scattered neutrons, so that the mPDF$(r)$ represents an ensemble average of quasi-instantaneous snapshots of the local {\it magnetic\/} structure, and thus includes both static and dynamic local spin correlations. In the case of a material that undergoes a long-range magnetic ordering transition, the spin correlations below the transition have a static {\color{black} (\ie~time-averaged)} component with a very long correlation length compared to the neutron coherence length. This produces elastic magnetic Bragg peaks whose integrated intensities can be used to refine a space- and time-averaged magnetic structural model, to which mPDF analysis then shows the spatially and temporally local deviations from that average magnetic structure.  Above the transition temperature, however, there are no magnetic Bragg peaks since the spin correlation lengths are very short, and the spin correlation times are also very short, such that there is essentially no elastic magnetic scattering that would lead to Bragg peaks. The local spin correlations are nevertheless accurately depicted as well-defined peaks in the mPDF$(r)$, since it is the Fourier transform of total-scattering magnetic diffraction intensity. This sensitivity to dynamic spin correlations above the magnetic transition temperature is one of the uniquely powerful aspects of mPDF analysis.



{\color{black}
The discussion above applies to powder samples. However, similar reasoning applies to single crystals, and the Fourier transform of diffuse magnetic scattering from a single crystal is called the \deltampdf~\cite{Roth+etal2018}, which has the advantage of resolving spin pairs via vectors in space, and not just by radial distance.
The ``$\Delta$'' refers to the prerequisite subtraction of the magnetic Bragg peak intensities, whose 3-dimensional Fourier transform would yield the magnetic Patterson function\cite{PattersonPR1934,PattersonZfK1935} representing the space- and time-average of the magnetic structure. If no magnetic Bragg peaks exist because the system does not exhibit long-range magnetic order, the magnetic Patterson function is zero. Either way, the \deltampdf\ represents the space- and time-local deviations from the magnetic Patterson function, and enables a similarly intuitive interpretation as did the powder mPDF: namely that positive and negative peaks in the \deltampdf\ correspond to excesses of parallel and antiparallel spin-pair alignments, respectively, as compared to the average magnetic structure.  
}

\subsection{Instruments and software for mPDF analysis}


Successful mPDF experiments require a neutron-scattering instrument with certain key characteristics. First, the instrument must have a sufficiently large $q$-range to measure enough of the magnetic scattering to enable a meaningful Fourier transform; in most cases, $q_{\rm min}$ should be around \color{black}0.5~\AA$^{-1}$~\color{black} or lower and $q_{\rm max}$ at least \color{black} 6 -- 8~\AA$^{-1}$,~\color{black} beyond which the magnetic form factor typically suppresses magnetic intensity too much to be measured accurately 
{\color{black}(but of course this depends on the magnetic species). 
}
We note that because of the magnetic form factor, the required $q_{\rm max}$ for mPDF is much smaller than that for traditional atomic PDF. Second, the instrument should have very low and stable background counts so that weak magnetic diffuse scattering intensity can be measured reliably. Third, the counting rate should be fast, as at a high-power reactor source or spallation source, so that good-quality data with low statistical noise can be collected within a reasonable time. Fourth, low-temperature sample environments such as a cryostat or closed-cycle refrigerator are often necessary, since magnetic transition temperatures are often well below room temperature. Finally, good $q$-resolution is also necessary if the scientific case requires sensitivity to magnetic correlations over long real-space length scales.

Neutron diffractometers optimized for traditional atomic PDF analysis, including those at both reactor and short-pulse spallation sources, often meet the above requirements and can therefore be used for mPDF analysis \color{black} without additional modification or calibration.~\color{black} Indeed, the majority of mPDF experiments published to date have been performed at traditional PDF instruments, often featuring simultaneous atomic and magnetic PDF analysis. Prominent examples include the hot-source neutron diffractometer D4c at Institut Laue-Langevin (ILL) \cite{Fischer+etal_D4c_2002, Fischer+etal_D4c_2000}, which is particularly valuable for its extremely low and stable background, NOMAD at the Spallation Neutron Source (SNS) \cite{Neuefeind+etal_NOMAD_2012}, which has a very high count rate, and NOVA at 
J-PARC\cite{Tsunoda+etal_NOVA_2023}. Other neutron total-scattering diffractometers such as GEM\cite{Hannon+etal_GEM_2005, Enderby+etal_GEM_2004} and  NIMROD\cite{Bowron+etal_NIMROD_2010} at the ISIS facility and POWGEN~\cite{huq;jac19} at SNS are also expected to be well suited for 
mPDF analysis.

Several diffractometers optimized for traditional, \ie~$q$-space refinement, magnetic and/or structural diffraction studies can also be used for mPDF analysis, as long as $q_{\mathrm{max}}$ is sufficiently high and the background is sufficiently low and stable. For these instruments, atomic PDF data cannot be collected simultaneously with mPDF data, since $q_{\mathrm{max}}$ is usually below 10~\AA$^{-1}$ and therefore not suitable for atomic PDF. Examples include D20 at ILL\cite{Hansen+etal_D4c_2008}, HB-2A at the High Flux Isotope Reactor~\cite{garle;apa10, baral;prb24}, and HYSPEC at SNS~\cite{zaliz;jpconfs17}. The latter instrument is notable for its use of polarized neutrons, which enables an automatic separation of atomic and magnetic scattering signals that can be useful for 
mPDF analysis~\cite{frand;jap22}. A more detailed discussion of advantages and disadvantages of reactor- versus spallation-based instruments and PDF versus $q$-space diffraction instruments in the context of mPDF analysis is warranted but will be deferred to a future article, together with additional notes on data collection and reduction procedures.

Once good-quality neutron scattering data have been collected, there remains the issue of how to isolate the desired magnetic diffraction intensity~$I_{\rm m}(q)$ from the sample's total diffraction intensity that also includes the ``nuclear scattering'' or scattering of neutrons from the atomic nuclei. One possibility is to collect data in the paramagnetic state well above any transition temperature and subtract this from data collected at lower temperatures, which can remove not only the magnetic self-scattering, but also the nuclear scattering in the case that the latter changed little between the two temperatures. The isolated $I_{\rm m}(q)$ can then be Fourier-transformed to mPDF$(r)$. Neutron-polarization methods may also be used to extract $I_{\rm m}(q)$. Alternatively, one can Fourier transform the total atomic and magnetic scattering simultaneously to obtain both the atomic and mPDF signals together in real space, and then model each component simultaneously using appropriate software. This approach, represented in Fig.~\ref{fig:schematic}, is often used at spallation-based PDF instruments such as NOMAD.

{\color{black}
Multiple open-source software packages exist that can facilitate mPDF analysis. The python package diffpy.mpdf\cite{Frandsen+etal_mpdf_2022} supports comprehensive mPDF functionality, including Fourier transforming magnetic total-scattering data to obtain the experimental mPDF$(r)$, simulating 1- and 3-dimensional mPDF patterns, and modeling measured mPDF data via least-squares refinement of 1-, 2-, or 3-dimensional unit-cell models. Additionally, simultaneous fits to the atomic PDF$(r)$ and mPDF$(r)$ can be performed with diffpy.mpdf when used in conjunction with the \texttt{diffpy-cmi} package~\cite{juhas;aca15}. Another option is Spinvert\cite{Paddison+etal_Spinvert_2013}, a popular and user-friendly program for producing a reverse Monte Carlo (RMC) model of spins for diffuse magnetic scattering data, which can then be Fourier transformed into a mPDF($r$) if desired, and also used to generate a useful spin-correlation function.
}
An advantage of fitting in $q$-space is that a spin configuration can be refined even when the experimental $q_{\rm max}$ is too low for Fourier transformation to a mPDF$(r)$. On the other hand, much of the intuitive power of mPDF analysis in real space is lost in $q$ space. The versatile RMCprofile\cite{Tucker+etal_RMCProfile_2007} software package is generally used to perform RMC modeling of nuclear scattering data, but it also has the option to include both Bragg and diffuse magnetic scattering. Such software has the potential to be quite valuable for mPDF analysis, notably due to its ability to separate the atomic and magnetic contributions to the measured total-scattering intensity through modeling rather than through temperature subtractions.  
{\color{black}
Of mention also is the powerful and user-friendly magnetic crystallography software package Mag2Pol\cite{Qureshi_Mag2Pol_2019} which treats powder and single-crystal data from polarized (and unpolarized) neutron diffractometers, and can also calculate the magnetic PDF$(r)$ for a given spin configuration.
}

\section{Recent applications of mPDF analysis to advance our understanding of emerging materials}

We now provide a few examples of mPDF analysis as applied to various types of magnetic materials that have recently emerged as important systems for technological applications and/or advancement of fundamental materials physics and chemistry. These examples, which include the antiferromagnetic semiconductor MnTe, selected geometrically frustrated magnets, and iron oxide magnetic nanoparticles, highlight the unique information that can be gained from mPDF analysis and further demonstrate the relevance of local magnetic structure to material properties. \color{black} Although the materials highlighted here could be considered unusual or remarkable in some way, local spin correlations often persist above the long-range magnetic ordering temperature even in more conventional magnets, which may therefore also benefit from mPDF analysis. \color{black} 

\subsection{MnTe: High-performance thermoelectricity and giant spontaneous magnetostriction}

\begin{figure}
    \centering
    \includegraphics[width=120mm]{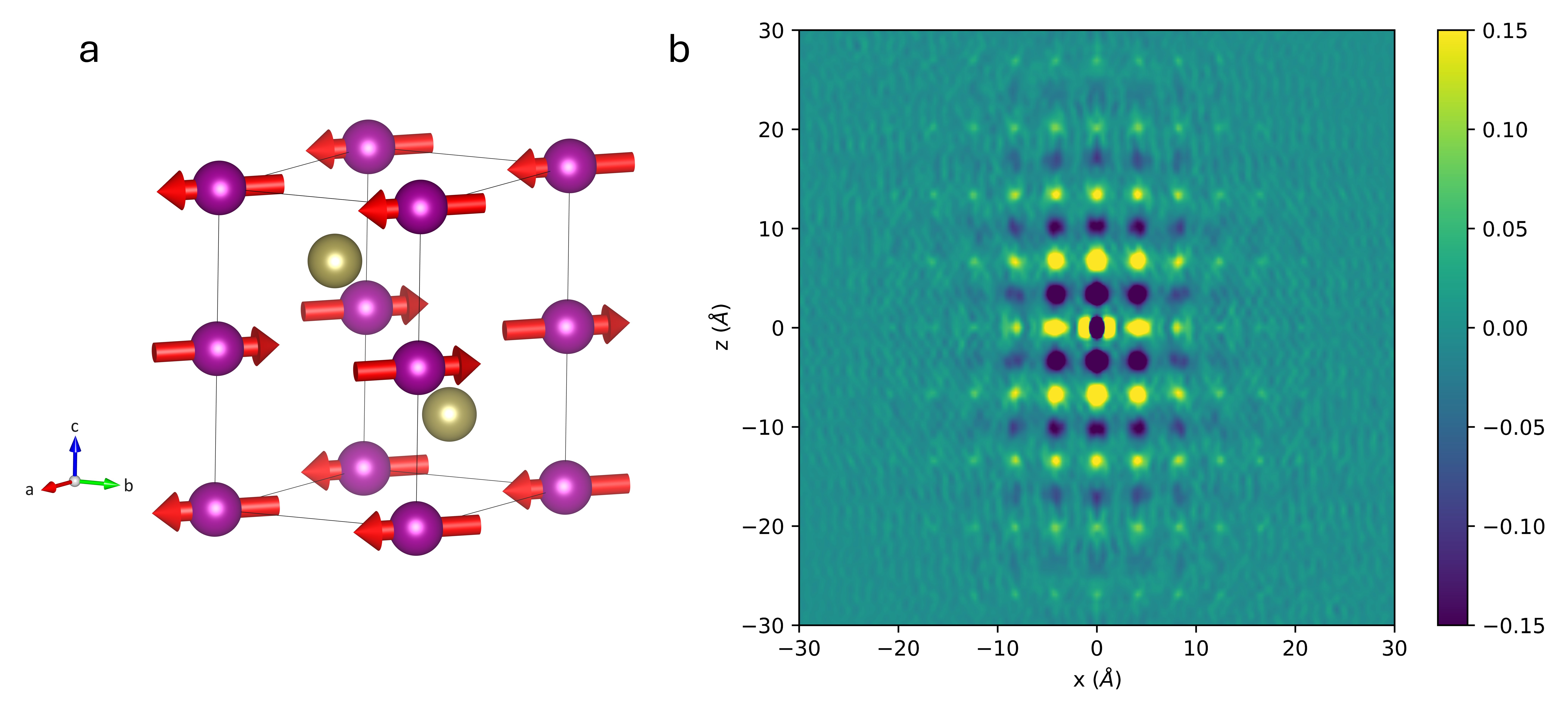}
    \caption{(a) Average atomic and magnetic structure of MnTe. (b) \deltampdf\ pattern for MnTe showing short-range, anisotropic AFM correlations in the \textit{xz} plane at $T \sim 340$~K. Positive (negative) values of the \deltampdf\ indicate net ferromagnetic (antiferromagnetic) correlations between spins separated by the corresponding real-space vector. Adapted with permission from Ref.~\citenum{baral;matter22}. Copyright 2022 Elsevier.}
    \label{fig:3DmPDF}
\end{figure}

MnTe is a hexagonal antiferromagnetic semiconductor with a N\'eel temperature of 307~K, below which the Mn$^{2+}$ ($S = 5/2$) moments align parallel to each other within the hexagonal \textit{ab} plane with moment directions that alternate along the \textit{c} axis [see Fig.~\ref{fig:3DmPDF}(a)]. Despite the rather simple average atomic and magnetic structure in MnTe, this material has recently attracted enormous attention in several different research areas for its remarkable and varied properties. It has been identified as a high-performance thermoelectric compound~\cite{ren;jmchemc17, xu;jmchema17, zulki;advs23}, a viable platform for antiferromagnetic spintronics~\cite{krieg;nc16}, a leading candidate for a material realization of the recently introduced notion of altermagnetism~\cite{mazin;prb23, kremp;n24}, and a magnetostructural material with an exceptionally large spontaneous magnetovolume effect~\cite{baral;afm23}. Magnetism underlies the exceptional behavior displayed by MnTe in all four of these areas of interest. Here, we focus on the role of short-range AFM correlations in driving the promising thermoelectric and magnetostructural responses in MnTe. 

Lightly doped samples of MnTe exhibit $zT$ in excess of 1 around 800~K, making MnTe a high-performance thermoelectric candidate~\cite{ren;jmchemc17, xu;jmchema17}. This large $zT$ is due primarily to an unusually high Seebeck coefficient that persists to elevated temperatures. Based on inelastic neutron scattering data and theoretical predictions~\cite{zheng;sadv19, polas;crps21}, this has been attributed to the influence of 
{\color{black}
paramagnons, which are dynamically correlated, short-range AFM spin fluctuations in the near paramagnetic state existing above the magnetic transition temperature. Essentially, paramagnons are like magnons (\ie~quantized spin waves) but shorter-lived and shorter-ranged because they exist in a quasi-paramagnetic state. Naturally, a perfectly paramagnetic state would have absolutely no correlations between distinct spins. 
}
Through a mechanism called ``paramagnon drag'', a temperature gradient induces a thermal flux of paramagnons, which in turn ``drag'' electrons through the lattice via exchange of linear momentum between the electrons and paramagnons. This mechanism is effective as long as the spatial and temporal extent of the paramagnons is sufficiently long, \ie~comparable to the effective Bohr radius and scattering time of mobile carriers.

An important contribution of mPDF was to probe these short-range AFM correlations directly to verify the paramagnon drag scenario~\cite{baral;matter22}. Fig.~\ref{fig:3DmPDF}(b) displays the
\deltampdf\ pattern generated from diffuse magnetic scattering data collected from a single crystal of MnTe at 340~K, above the N\'eel temperature of 307~K.
Light and dark spots occur at interatomic vectors corresponding to pairs of Mn atoms whose spins are parallel and antiparallel, respectively, with \textit{x} and \textit{z} indicating interatomic vector components along the  crystallographic \textit{a} and \textit{c} directions. The \deltampdf\ pattern therefore confirms the presence of well-defined short-range AFM correlations on the length scale of a few nanometers in the otherwise paramagnetic state of MnTe. Furthermore, the rows of light and dark spots that alternate along \textit{z} reveal a short-range-ordered version of the average AFM structure, with parallel alignment of spins within the \textit{ab} plane and antiparallel alignment between neighboring spins along \textit{c}. Interestingly, the correlation length is about twice as long along \textit{z} as it is along \textit{x}, which can be explained by the much stronger magnetic exchange interaction between nearest-neighbor spins along \textit{c} than among the neighboring spins in the hexagonal \textit{ab} plane.

\begin{figure*}
    \centering
    \includegraphics[width=120mm]{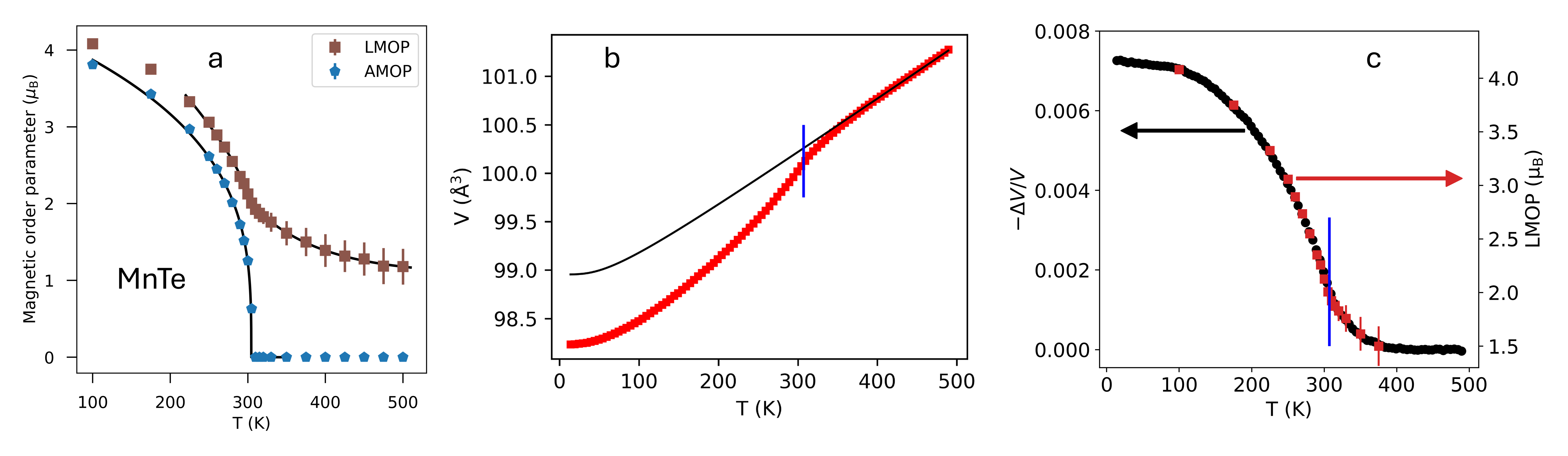}
    \caption{ (a) Temperature dependence of the average magnetic order parameter (AMOP) and local magnetic order paramater (LMOP) as determined by mPDF fits. The LMOP corresponds to the instantaneous AFM correlations between nearest-neighbor spins. (b) Unit cell volume versus temperature obtained from x-ray PDF data. The blue vertical line marks \TN\ at the point of inflection, and the black curve indicates the expected unit cell volume if magnetism were absent in MnTe. (c) Comparison of the magnetically-induced fractional volume change $-\Delta V/V$ (black circles, left vertical axis), to the LMOP (red squares, right vertical axis). Adapted with permission from Ref.~\citenum{baral;afm23}. Copyright 2023 John Wiley and Sons.}
    \label{fig:MnTe-Tdep}
\end{figure*}

Temperature-dependent mPDF data collected on a powder sample of MnTe demonstrates that the short-range AFM correlations are remarkably robust with temperature, again supporting the paramagnon drag picture. This can be seen in Fig.~\ref{fig:MnTe-Tdep}(a), where the local magnetic order parameter (LMOP, \color{black} \ie~the magnitude of the instantaneously correlated component of nearest-neighbor spins expressed in Bohr magnetons\color{black}) extracted from mPDF fits shows only a slow decrease with temperature above \TN, remaining above 1~$\mu_{\mathrm{B}}$ at 500~K, thus \color{black}demonstrating the robustness of instantaneous short-range spin correlations above \TN\ and~\color{black} supporting the presence of paramagnons at high temperatures.
\color{black}The correlation length (not shown) has a relatively constant value of several Angstroms above 400~K and grows steadily as the temperature decreases below 400~K until long-range order is achieved at \TN.~\color{black} The behavior of the LMOP contrasts with that of the average magnetic order parameter (AMOP, \ie~the traditional ``ordered moment'' as averaged over space and time), which shows the typical decrease to zero upon warming through \TN. This underscores how the local and average magnetic structures can be dramatically different in a large temperature window above the long-range ordering temperature. 

Magnetic PDF analysis has also proven useful for understanding magnetostructural coupling in MnTe~\cite{baral;afm23}. Fig.~\ref{fig:MnTe-Tdep}(b) shows the unit cell volume $V$ of MnTe determined from fits to the atomic PDF obtained from x-ray scattering data. The fits were performed over the data range 0 -- 50~\AA, reflecting the structure on this length scale. Upon cooling from high temperature, the experimentally determined volume (red symbols) initially shows the typical linear thermal contraction, but then drops below the high-temperature trend starting around 400~K. This anomalous contraction of the unit cell volume is most pronounced around $T_{\mathrm{N}}=307$~K, indicating that the magnetism in MnTe is driving the lattice contraction and is thus a manifestation of the spontaneous magnetovolume effect. The expected temperature dependence of the unit cell volume in the absence of any magnetic effects is given by the black curve in Fig.~\ref{fig:MnTe-Tdep}(b), which was calculated using the Debye-Gr\"uneisen model~\cite{baral;afm23}. The fact that the lattice contraction begins well above \TN\ strongly suggests that the short-range AFM correlations in the otherwise paramagnetic state begin driving the global lattice response even before the onset of long-range AFM order. This is confirmed by overlaying the LMOP from Fig.~\ref{fig:MnTe-Tdep}(a) on the fractional volume change $-\Delta V/V$, where $\Delta V$ is the difference between the expected (nonmagnetic) unit cell volume and the observed volume, as displayed in Fig.~\ref{fig:MnTe-Tdep}(c). A clear one-to-one correspondence is seen even above \TN, confirming that the local magnetic structure (whether above or below \TN) is responsible for the change in unit cell volume, once again highlighting the strong influence that short-range magnetism can have on material properties. These results also demonstrate that short-range correlations in one degree of freedom (spin) can couple efficiently to longer-range correlations in another degree of freedom (lattice). Incidentally, the unit cell contraction in MnTe reaches a remarkably large value of nearly 1\% at low temperature, qualifying as the largest known spontaneous magnetovolume effect in an antiferromagnet~\cite{baral;afm23}.

\subsection{Geometrically frustrated magnets: Insights from static and dynamic short-range spin correlations revealed by mPDF analysis}

The sensitivity of mPDF to short-range correlations makes it an especially valuable tool for studying geometrically frustrated magnets. As described previously, frustration inhibits the formation of long-range order, such that the system orders at anomalously low temperatures or not at all, as for a QSL. In either case, short-range correlations (either static or dynamic) are ubiquitous in frustrated magnets and contain vital information about the underlying physics of the material, making them a prime candidate for mPDF analysis.

\begin{figure*}
    \centering
\includegraphics[width=120mm]
  {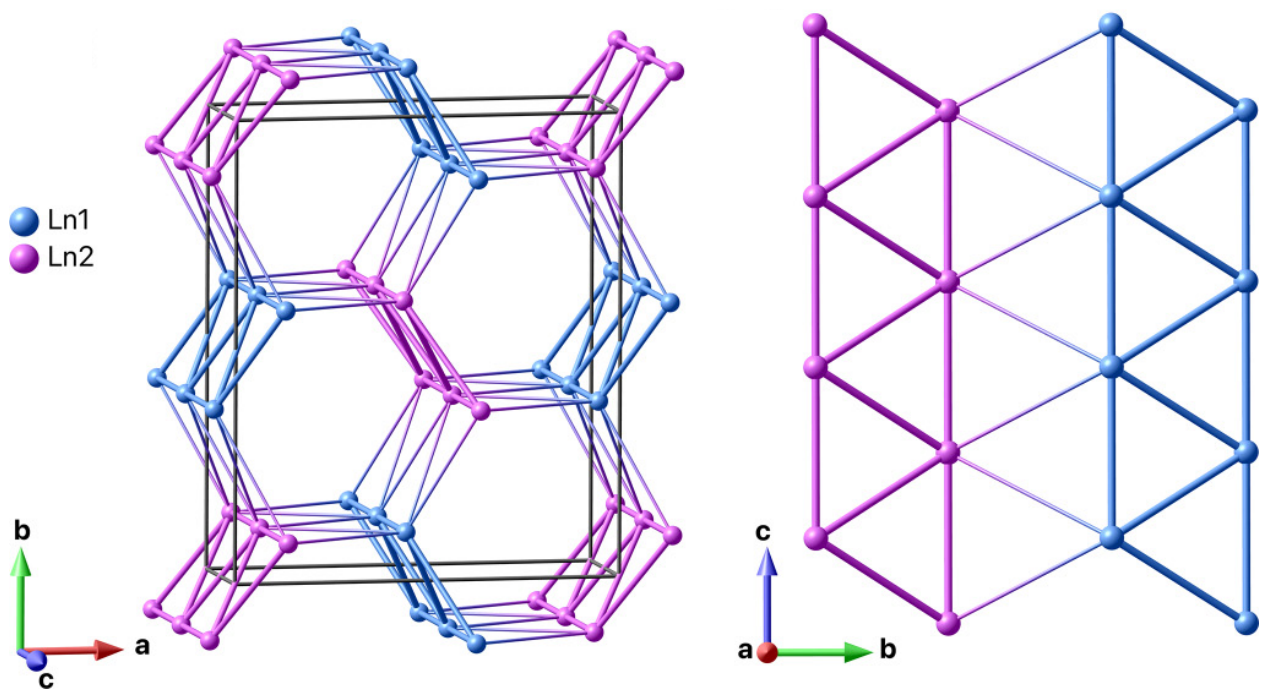}
    \caption{The Ln magnetic ions of the \ce{SrLn2O4} structure form a distorted honeycomb lattice (left) whose hexagonal sheets are linked along the $c$-axis by two slightly different types of zig-zag ``ladders'' (right) wherein NN ions reside along the same ladder pole, and NNN ions are connected by diagonal ``rungs'' within a given ladder.  The triangular geometry of the zig-zag ladders gives rise to magnetic frustration for AFM interactions.  Reproduced with permission from Ref.~\citenum{Qureshi+etal2022}. Copyright 2022 American Physical Society.}
    \label{fig:Qureshi_structure}
\end{figure*}

The study by Qureshi, \etal,\cite{Qureshi+etal2022} on the frustrated magnetism of lanthanide strontium oxides provides an excellent example of the power of magnetic PDF analysis to follow the temperature evolution of short-range spin correlations continuously as the magnetic transition temperature is crossed.  As shown in the left side of Fig.~\ref{fig:Qureshi_structure}, the magnetic ions of the \ce{SrLn2O4} structure form a distorted honeycomb lattice wherein the hexagonal sheets are stacked along the c-axis and connected by zig-zag ``ladders'', shown in detail at the right.  Each magnetic ion has a nearest neighbor (NN) along the same pole of the ladder, and then two next-nearest neighbors (NNN) on the opposite pole as linked by diagonal ``rungs''.  These neighboring ions form ``intra-ladder'' spin pairs, as opposed to ``inter-ladder'' pairs at greater distances.  Antiferromagnetic (AFM) interactions between NN and NNN are geometrically frustrated by the triangular geometry of their zig-zag ladder.  Note that there are two different ladders of Ln sites having slightly different dimensions and crystallographic environments for their magnetic ions.

\begin{figure*}
    \centering
\includegraphics[width=160mm]
  {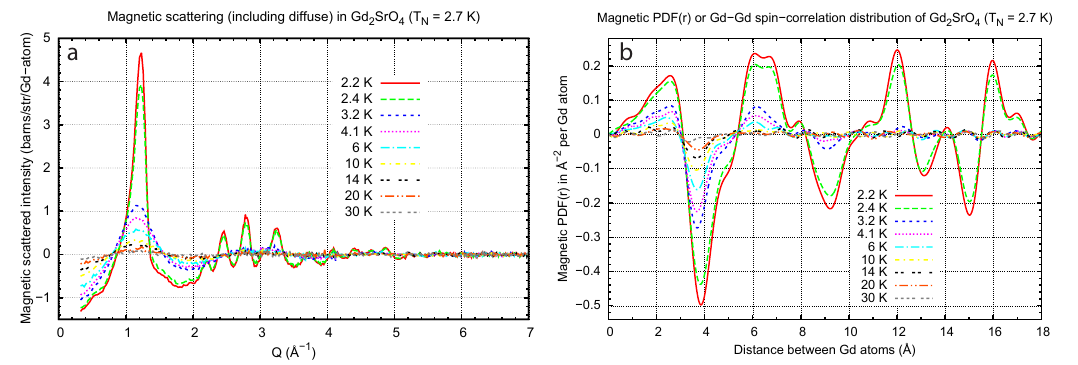}
    \caption{Total magnetic scattering from \ce{SrGd2O4} in $q$-space (a) and $r$-space (b) after subtraction of the paramagnetic contribution at 50~K representing uncorrelated Gd$^{3+}$ spins. The normalized mPDF$(r)$ were calculated from the experimental data using the first line of Eq.~\ref{eqn:mPDF_<F>}. Adapted with permission from Ref.~\citenum{Qureshi+etal2022}. Copyright 2022 American Physical Society.}
    \label{fig:Qureshi_q+r_space}
\end{figure*}

The \ce{SrGd2O4} compound ($T_{\rm N}\approx 2.73$~K) is particularly illustrative.  Figure~\ref{fig:Qureshi_q+r_space}(a) shows the total measured magnetic diffraction intensity resulting from Gd$^{3+}$ spin correlations at several temperatures, which was obtained by subtracting from the total-scattering diffraction pattern of each temperature the diffraction pattern at 50~K as a paramagnetic ``baseline'' comprising nearly uncorrelated spins.  Such a subtraction also eliminates the nuclear diffraction intensity as long as there is little thermal contraction of the lattice below 50~K, as was the case here.  The magnetic scattering intensity was then Fourier transformed to produce the mPDF$(r)$ of Fig.~\ref{fig:Qureshi_q+r_space}(b), where weak AFM spin correlations are still visible at the Gd-Gd next-nearest-neighbor distance of $\sim 3.6$~\AA\ for temperatures up to 10 times $T_{\rm N}$.

\begin{figure*}
    \centering
\includegraphics[width=160mm]
  {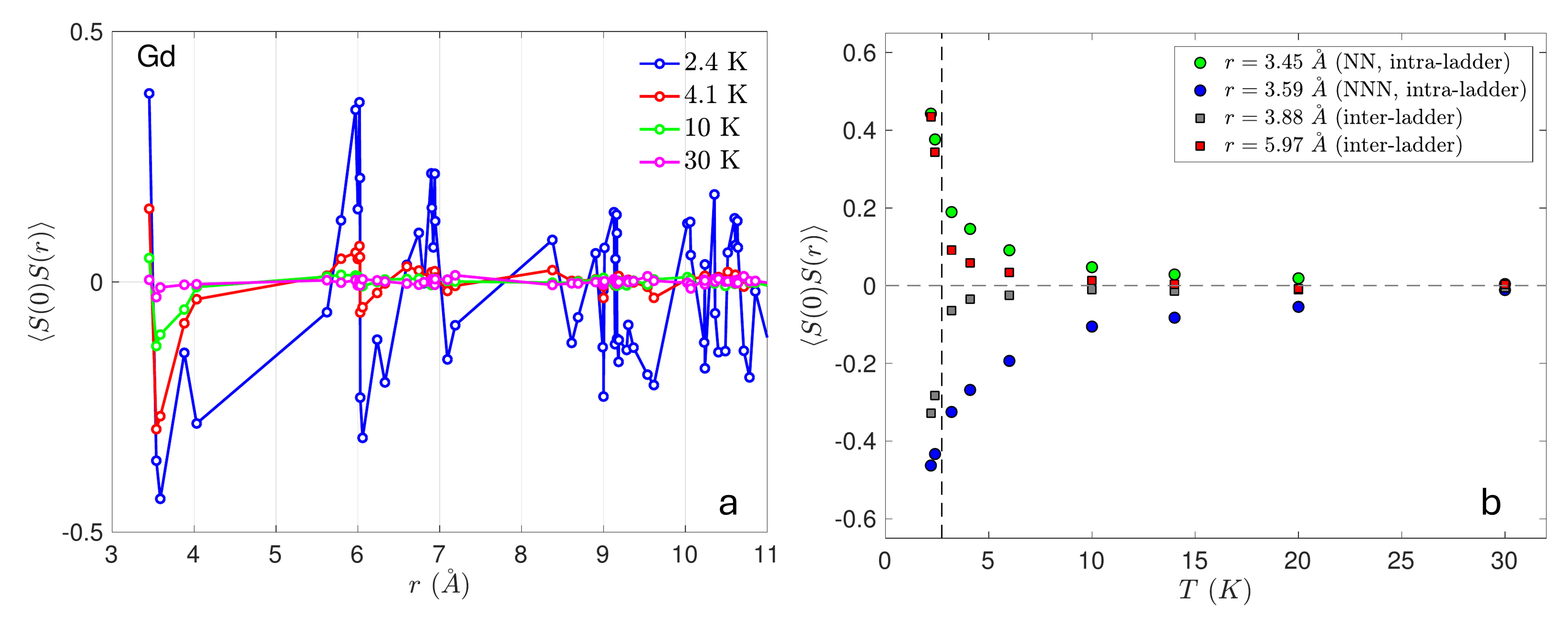}
    \caption{Spin correlations in \ce{SrGd2O4} between different spins pairs at a given temperature (a) as a function of inter-spin distance, and the temperature dependence of the spin correlation for selected spin pairs (b) both within and between ``zig-zag ladders'' of the structure.  Adapted with permission from Ref.~\citenum{Qureshi+etal2022}. Copyright 2022 American Physical Society.}
    \label{fig:Qureshi_SCF_r+T_dependence}
\end{figure*}

Most striking in the mPDF$(r)$ curves is the contrast between a nearly continuous increase in spin correlations at the shortest inter-spin distances as the magnetic transition is crossed, as compared to a sudden jump of mPDF amplitude at larger inter-spin distances.  This temperature dependence of spin pair correlations as a function of spin separation can be made site-specific by performing RMC simulations using Spinvert\cite{Paddison+etal_Spinvert_2013}, and then calculating the spin correlation function (SCF) which is simply the dot product of relative spin orientations for a given pair of spin sites as averaged over all such spin pairs.  Figure~\ref{fig:Qureshi_SCF_r+T_dependence}(a) shows the resulting SCF$(r)$ function for one temperature below $T_{\rm N}$ and three above.  The increasing discontinuity in spin correlations at $T_{\rm N}$ with increasing inter-spin distance is made even more clear in Fig.~\ref{fig:Qureshi_SCF_r+T_dependence}(b) showing the temperature dependence of the SCF for two intra-ladder spin pairs (green and blue circles), as compared to that for two inter-ladder spin pairs (grey and red squares).  As the sample is cooled toward $T_{\rm N}$ therefore, the intra-ladder correlations begin ordering very soon due to the strong NN and NNN interactions, whereas the inter-ladder correlations become significant only slightly above $T_{\rm N}$.
The take-home message from this study is that short-range dynamic spin correlations can persist up to 10 times $T_{\rm N}$, at least in a frustrated magnetic system, even if long-range dynamic spin correlations have vanished soon above $T_{\rm N}$, and that there is a continuous increase in the abruptness of the onset of spin correlations at $T_{\rm N}$ with increasing inter-spin distance.

\begin{figure*}
    \centering
    \includegraphics[width=160mm]{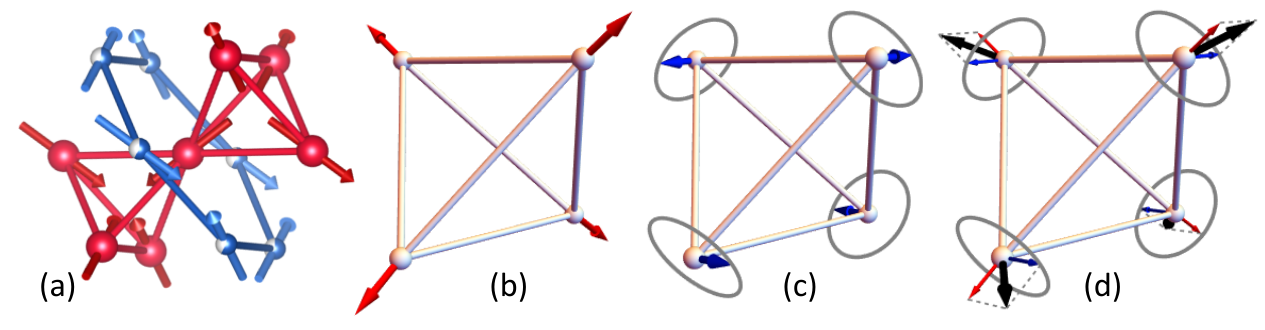}
    \caption{(a,b) Expected AIAO configuration of the Gd spins (red) in \ce{Gd2Ir2O7} as induced by the AIAO-ordered Ir spins (blue) and as obtained from Rietveld refinement but with a too small ordered Gd$^{3+}$ moment of $M_{||} = 4.35~\mu_{\rm B}$ as compared to the nominal 7.94~$\mu_{\rm B}$. (c) The Palmer-Chalker spin configuration, and (d) a coexistence of AIAO and Palmer-Chalker configurations. Reproduced with permission from Ref.~\citenum{Lefrancois+etal2019}. Copyright 2019 American Physical Society.}
    \label{fig:Lefrancois_AIAO+PC}
\end{figure*}

Other than determining quantitatively the short-range spin correlations both below and above the magnetic transition temperature, mPDF analysis can also reveal additional complexity in an ordered magnetic system that could not be deduced from conventional Rietveld refinement of magnetic Bragg peaks.  For instance, in the case of the pyrochlore \ce{Gd2Ir2O7}, both the Gd and Ir spins are geometrically frustrated, as they each populate the vertices of separate pyrochlore sublattices.  The Ir spins order into an All-In-All-Out (AIAO) configuration at 120~K, and it was expected that the Gd spins do likewise on their sublattice at much lower temperature (see Figs.~\ref{fig:Lefrancois_AIAO+PC}(a) and (b)).  Indeed, as reported by Lefran\c{c}ois, \etal,\cite{Lefrancois+etal2019} Rietveld refinement of magnetic Bragg peak intensities from neutron powder diffraction data taken at 1.4~K confirmed an AIAO configuration for the Gd spins, but with an ordered magnetic moment of only $M_{||} = 4.35~\mu_{\rm B}$ as compared the full moment of 7.94~$\mu_{\rm B}$ for Gd$^{3+}$ spins.  The apparent spin disorder was consistent with the observation of a broad bump of magnetic diffuse scattering at $q\sim 1.2$~\AA$^{-1}$.

\begin{figure*}
    \centering
\includegraphics[width=160mm]
  {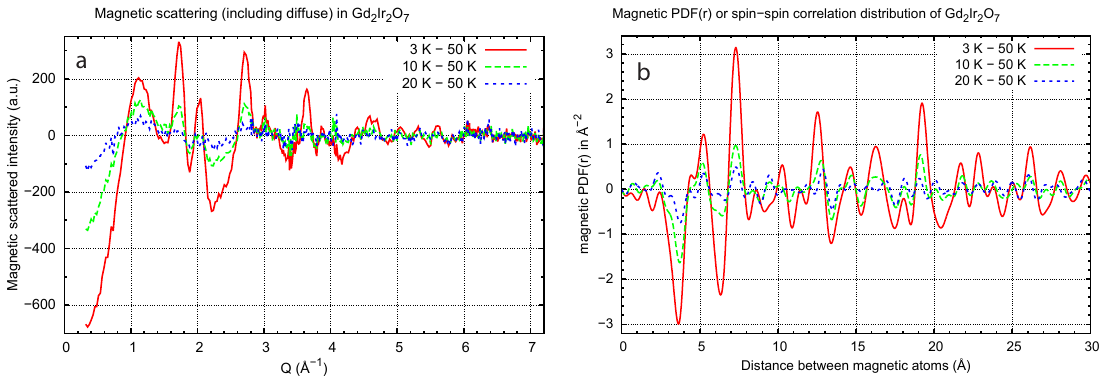}
    \caption{Total magnetic scattering from the \ce{Gd2Ir2O7} pyrochlore in $q$-space (a) and $r$-space (b) after subtraction of the paramagnetic contribution at 50~K representing uncorrelated Gd$^{3+}$ spins. Note the the strong AFM correlation at the nearest-neighbor Gd-Gd distance of $r\sim 3.6$~\AA, corresponding to one edge of a Gd tetrahedron.  The normalized mPDF$(r)$ were calculated from the experimental data using the first line of Eq.~\ref{eqn:mPDF_<F>}. Adapted with permission from Ref.~\citenum{Lefrancois+etal2019}. Copyright 2019 American Physical Society.}
    \label{fig:Lefrancois_q+r_space}
\end{figure*}

An mPDF experiment was then performed as a function of temperature, where again the total magnetic scattering intensity due to Gd$^{3+}$ spin correlations was isolated from the nuclear scattering via the subtraction of a 50~K paramagnetic ``baseline'' diffraction pattern from the lower temperature patterns, and then Fourier transformed to produce mPDF$(r)$ for temperatures of 3~K, 10~K and 20~K (see Fig.~\ref{fig:Lefrancois_q+r_space}).  It was assumed that the Ir spin configuration did not change between 50~K and 3~K.

\begin{figure*}
    \centering
\includegraphics[width=160mm]
  {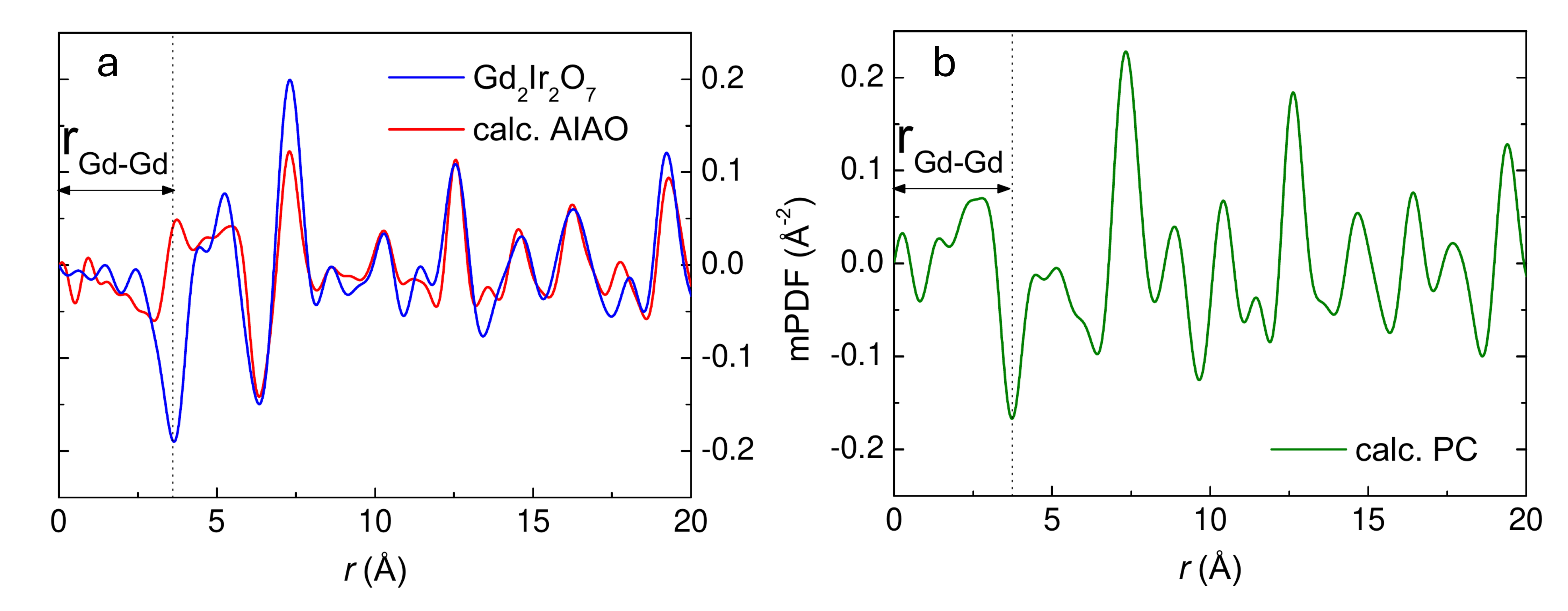}
    \caption{(a): mPDF($r$) at 3~K for the Gd$^{3+}$ spins as obtained from a Fourier transform of the data in Fig.~\ref{fig:Lefrancois_q+r_space}(a), and as calculated for the expected AIAO spin configuration of Fig.~\ref{fig:Lefrancois_AIAO+PC}(b).  (b): mPDF($r$) as calculated for the Palmer Chalker (PC) spin configuration of Fig.~\ref{fig:Lefrancois_AIAO+PC}(c), which reproduces the AFM correlations found experimentally at $r\sim 3.6$~\AA.  Reproduced with permission from Ref.~\citenum{Lefrancois+etal2019}. Copyright 2019 American Physical Society.}
    \label{fig:Lefrancois_mPDF_data_vs_models}
\end{figure*}

A comparison of the measured mPDF$(r)$ with that calculated for an AIAO model of Gd$^{3+}$ spins showed a significant qualitative discrepancy in the short-range spin correlations as shown in Fig.~\ref{fig:Lefrancois_mPDF_data_vs_models}(a), namely that the AIAO model predicted a weak ferromagnetic (FM) correlation between nearest-neighbor Gd-Gd spins at $r\sim 3.6$~\AA, whereas the data clearly showed a strong antiferromagnetic (AFM) correlation between the same spins.  Such a discrepancy between model and data was not at all evident from the conventional Rietveld refinement that had been carried out.  Additional modeling then showed that the Gd-Gd nearest-neighbor AFM correlations could be reproduced (see Fig.~\ref{fig:Lefrancois_mPDF_data_vs_models}(b)) via a coexistence of Palmer Chalker (PC) spin correlations between Gd spins in addition to the AIAO correlations, as illustrated in Figs.~\ref{fig:Lefrancois_AIAO+PC}(c) and (d).  The observed PC correlations showed that it was necessary to take into account a weak easy-plane Gd$^{3+}$ anisotropy in the context of the  staggered molecular field induced by the AIAO ordering of the Ir ions, instead of making the usual assumption that Gd-Gd interactions can be treated as isotropic 3D-Heisenberg.  Such an insight and the resulting determination of the low-temperature magnetic structure of \ce{Gd2Ir2O7} might never have been made without magnetic PDF analysis, since it reveals short-range spin correlations quantitatively and independently of any modeling bias.

\begin{figure*}
    \centering
    \includegraphics[width=160mm]{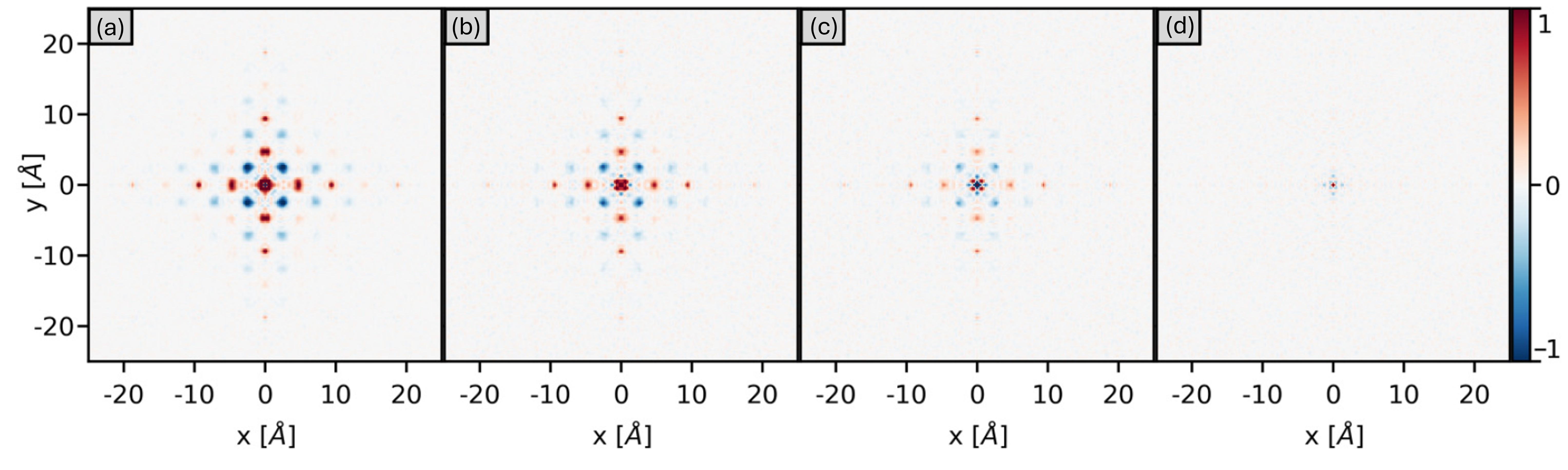}
    \caption{ 3D-$\Delta$mPDF patterns for natural spin glass mineral bixbyite with composition Fe$_{1.12}$Mn$_{0.88}$O$_3$ at (a) 7~K, (b) 50~K, (c) 80~K, and (d) 240~K. The spin-glass freezing temperature is 32~K. The patterns correspond to the $z=0$ plane. Reproduced with permission from Ref.~\citenum{Roth+etal2019}. Copyright 2019 American Physical Society.}
    \label{fig:bixbyite}
\end{figure*}

The naturally occurring mineral bixbyite with the chemical formula \ce{(Mn_{1-x}Fe_x)2O3} showcases the value of \deltampdf\ for studying frustrated magnets. The Mn and Fe atoms are randomly distributed on shared crystallographic sites that form triangles and hexagons within the overall cubic structure. This combination of substitutional disorder and frustration gives rise to a spin-glass transition, in which the spins freeze into a short-range-ordered configuration below the spin-glass freezing temperature. Spin glasses have attracted enormous attention over the past several decades for their highly unusual magnetic properties and their role as model systems for complexity~\cite{mydos;rpp15}. Roth \etal~\cite{Roth+etal2018, Roth+etal2019} performed \deltampdf\ analysis on a single crystal of Fe$_{1.12}$Mn$_{0.88}$O$_3$, which has a spin-glass transition temperature of 32~K, to probe the local instantaneous spin correlations.  The \deltampdf\ patterns are shown at various temperatures above and below the transition temperature in Fig.~\ref{fig:bixbyite}.
The pattern at 7~K reveals clear short-range AFM correlations with a length scale of $\sim$15~\AA, confirming the disordered spin-glass ground state. Interestingly, the local instantaneous AFM configuration is qualitatively unchanged \textit{above} the spin-glass transition up to about 240~K, aside from an overall weaker amplitude due to increased thermal disorder. This indicates that the local magnetic environment characteristic of the frozen spin-glass state is preserved in the otherwise paramagnetic state, but as dynamically fluctuating spin correlations. The ability to probe this behavior directly and quantitatively is fairly unique to mPDF analysis and sheds new light on the decades-old spin glass phenomenon.

Magnetic PDF has been successfully applied to even more exotic geometrically frustrated magnets, such as candidate QSLs~\cite{Nuttall+etal2023} and a material realizing a topological Kosterlitz-Thouless phase~\cite{Dun+etal2021}, which we will not discuss in detail here. In all cases, the ability to visualize and model the local magnetic correlations through mPDF analysis has resulted in substantially deeper understanding of the properties of each specific system as well as the general physics of geometrically frustrated magnets.

\subsection{Magnetic nanoparticles}

Nanosized particles of ferro- or ferrimagnetic compounds such as the iron-oxide spinels \ce{Fe3O4} (magnetite) and $\gamma$-\ce{Fe2O3} (maghemite) have recently seen an explosion of interest due to their applications in diverse fields ranging from medicine to information storage~\cite{singa;jmc11, majet;acsn11, mater;assa21, kianf;jsnm21}. For example, they can be used for targeted treatment of cancerous tumors through the phenomenon of magnetic hyperthermia~\cite{fatim;nanom21}, by which an alternating magnetic field excites magnetic nanoparticles that have been delivered to the tumor, increasing the temperature locally to destroy the tumor with minimal damage to surrounding tissues. Additionally, the superparamagnetic properties of magnetic nanoparticles have proven to be extremely effective for enhancing contrast in magnetic resonance imaging (MRI), thus improving its diagnostic capabilities~\cite{na;am09}.

Despite the broad applications of iron-oxide nanoparticles that are already in advanced stages of development, several basic aspects of their chemistry, structure, and magnetism have remained unclear for many years. By definition, nanoparticles have no long-range crystal structure, so PDF methods provide a crucial alternative to conventional Bragg diffraction for investigating their atomic and magnetic structure. A recent study of iron-oxide spinel nanoparticles provided the first use of mPDF analysis to investigate magnetic nanoparticles~\cite{ander;ij21}. In this study, a novel surfactant-free, hydrothermal synthesis method was used to produce gram-scale samples of iron-oxide nanoparticles in dried powder form with average particle sizes ranging from approximately 8~nm to 25~nm. Several structural probes, including x-ray atomic PDF, were used to ascertain that the composition was intermediate between \ce{Fe2O3} and \ce{Fe3O4}. The best structural model was found to be a modified version of the tetragonal structure of $\gamma$-\ce{Fe2O3}, in which Fe vacancies partially order into a superstructure. Interestingly, the structural correlation length of the vacancy superstructure was shorter than the overall particle size, indicating a non-uniform local structure that is nevertheless distinct from the core-shell nanoparticle picture that had been suggested previously~\cite{friso;cm13}. This implies a compositional gradient across the nanoparticles, with the Fe-rich vacancy-ordered center gradually evolving into a more oxidized and disordered outer region.

\begin{figure}
    \centering
    \includegraphics[width=120mm]{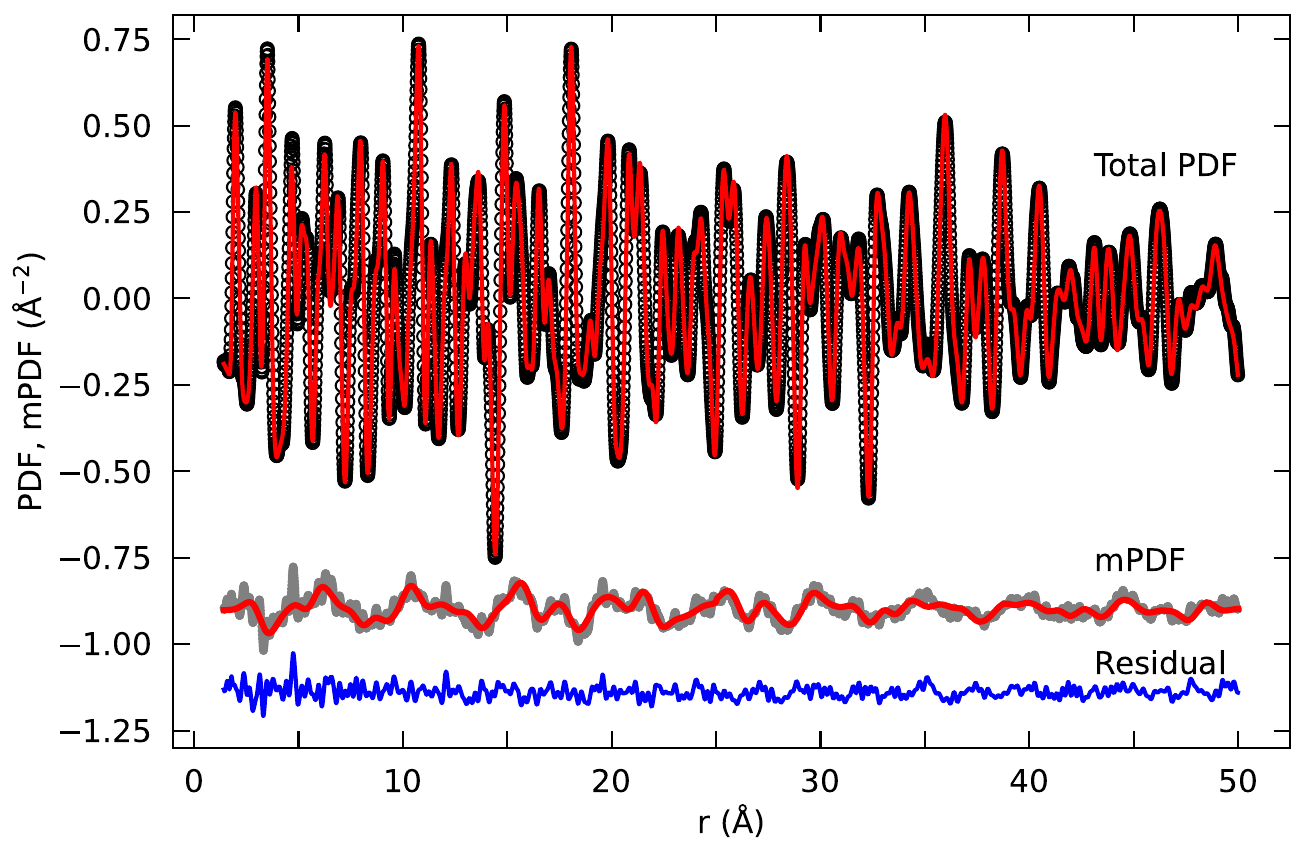}
    \caption{ Atomic and magnetic PDF analysis of iron-oxide spinel nanoparticles. Descriptions of the fits and how the mPDF signal was isolated are given in the main text. Adapted from Ref.~\citenum{ander;ij21}. Available under a CC-BY license. Copyright 2021 Henrik L. Anderson \etal}
    \label{fig:magNPs}
\end{figure}

Magnetic PDF was used to provide the missing information about the local magnetic structure. Combined atomic and magnetic PDF fits were performed using neutron total-scattering data collected from NOMAD at the SNS. The modified $\gamma$-\ce{Fe2O3} model used for the atomic x-ray PDF fits described in the previous paragraph was again used for the atomic PDF in the neutron data, while the ferrimagnetic structure previously observed in bulk \ce{Fe3O4} and $\gamma$-\ce{Fe2O3} was used as the starting magnetic model for the mPDF component. A representative fit is shown in Fig.~\ref{fig:magNPs}.
In this figure, the upper set of black and red curves represent the experimental and best-fit total PDF patterns, respectively, which include both the atomic and magnetic PDF components. Offset vertically below, the overlaid gray and red curves represent the isolated experimental and best-fit mPDF components, respectively. The experimental mPDF was obtained by subtracting the calculated atomic PDF from the total PDF data. The overall fit residual shown in blue is relatively small and flat, confirming the good quality of the fit.

Two aspects of the mPDF analysis are noteworthy. First, these results demonstrate that mPDF can be successfully applied to magnetic nanoparticles, opening the door to similar studies in the future. Second, the magnetic correlation length determined from the fits was found to be 15 $\pm$ 2~nm, significantly shorter than the overall particle size of 24 $\pm$ 1~nm determined from the atomic PDF fit. This is evident in the figure from the more rapid damping in real space of the mPDF compared to the total PDF, which is dominated by the atomic PDF. This provides an important and otherwise unknowable insight into the nature of these iron-oxide nanoparticles, namely that the magnetic structure does not extend uniformly throughout the entire nanoparticle. This could be due to deviations from the ferrimagnetic order near the surface of the particles because of spin canting, for example.

\section{Conclusion and Outlook}

In this Perspective, we have demonstrated the importance of local magnetic structure for the physical properties and fundamental physics of a wide variety of materials that have attracted recent fundamental and applied interest. We have additionally introduced mPDF analysis as a powerful probe of both static and dynamic short-ranged magnetic spin correlations via the Fourier transform of diffuse magnetic scattering data obtained from neutron total-scattering diffraction experiments. By transforming the diffuse scattering from reciprocal space into real space, short-range spin correlations both above and below magnetic phase transitions can be visualized directly, understood intuitively, and modeled quantitatively, offering a uniquely detailed view into local magnetic structure. 

As a relatively young experimental technique first formalized just 10 years ago, mPDF analysis offers several exciting avenues for further development that will enable new science. For example, the use of polarized neutrons is expected to improve significantly the quality and real-space clarity of mPDF data, providing greater sensitivity to subtle or complex features of local magnetic structure~\cite{frand;jap22}. In analogy to dynamic atomic PDF analysis of energy-resolved inelastic neutron scattering data~\cite{kimbe;nm23}, recent work has demonstrated the viability of dynamic mPDF analysis~\cite{Iida+etal2022}, which provides unique access to energy- and time-dependent features of short-range spin correlations and will likely be valuable for studying excitations in systems such as exotic quantum magnets. Working toward more routine \deltampdf\ analysis through improved data reduction and modeling techniques will enable an unprecedented level of detail in studies of short-range magnetism in crystals. Furthermore, next-generation neutron instruments designed with mPDF analysis in mind are set to come online in the coming years at the European Spallation Source (DREAM), Second Target Station of the Spallation Neutron Source (VERDI), and potentially elsewhere, which will open up even more doors for transformative mPDF developments and applications. Finally, even with the instruments currently available, we envision high-impact applications of mPDF to numerous material systems not discussed at length here, including magnetic topological materials, high-entropy magnetic materials, strongly correlated electron systems, molecular magnets, and more. The future is bright for mPDF analysis and the study of local magnetic structure in emerging materials.

\begin{acknowledgement}
We thank Simon Billinge for valuable discussions and long-time collaboration on magnetic pair distribution function methods. 
We are also very grateful for helpful discussions with Andrew Wildes, Navid Qureshi, Joe Paddison and Gerry Lander.  
B.A.F. was supported by the U.S. Department of Energy, Office of Science, Basic Energy Sciences (DOE-BES) through Award No. DE-SC0021134.

\end{acknowledgement}






\begin{mcitethebibliography}{121}
	\providecommand*\natexlab[1]{#1}
	\providecommand*\mciteSetBstSublistMode[1]{}
	\providecommand*\mciteSetBstMaxWidthForm[2]{}
	\providecommand*\mciteBstWouldAddEndPuncttrue
	{\def\EndOfBibitem{\unskip.}}
	\providecommand*\mciteBstWouldAddEndPunctfalse
	{\let\EndOfBibitem\relax}
	\providecommand*\mciteSetBstMidEndSepPunct[3]{}
	\providecommand*\mciteSetBstSublistLabelBeginEnd[3]{}
	\providecommand*\EndOfBibitem{}
	\mciteSetBstSublistMode{f}
	\mciteSetBstMaxWidthForm{subitem}{(\alph{mcitesubitemcount})}
	\mciteSetBstSublistLabelBeginEnd
	{\mcitemaxwidthsubitemform\space}
	{\relax}
	{\relax}
	
	\bibitem[Billinge and Kanatzidis(2004)Billinge, and Kanatzidis]{billi;cc04}
	Billinge,~S. J.~L.; Kanatzidis,~M.~G. Beyond crystallography: the study of disorder, nanocrystallinity and crystallographically challenged materials. \emph{Chem. Commun.} \textbf{2004}, \emph{7}, 749--760\relax
	\mciteBstWouldAddEndPuncttrue
	\mciteSetBstMidEndSepPunct{\mcitedefaultmidpunct}
	{\mcitedefaultendpunct}{\mcitedefaultseppunct}\relax
	\EndOfBibitem
	\bibitem[Dagotto(2005)]{dagot;s05}
	Dagotto,~E. Complexity in strongly correlated electronic systems. \emph{Science} \textbf{2005}, \emph{309}, 257--262\relax
	\mciteBstWouldAddEndPuncttrue
	\mciteSetBstMidEndSepPunct{\mcitedefaultmidpunct}
	{\mcitedefaultendpunct}{\mcitedefaultseppunct}\relax
	\EndOfBibitem
	\bibitem[Young and Goodwin(2011)Young, and Goodwin]{young;jmc11}
	Young,~C.~A.; Goodwin,~A.~L. Applications of pair distribution function methods to contemporary problems in materials chemistry. \emph{J. Mater. Chem.} \textbf{2011}, \emph{21}, 6464--6476\relax
	\mciteBstWouldAddEndPuncttrue
	\mciteSetBstMidEndSepPunct{\mcitedefaultmidpunct}
	{\mcitedefaultendpunct}{\mcitedefaultseppunct}\relax
	\EndOfBibitem
	\bibitem[Keen and Goodwin(2015)Keen, and Goodwin]{keen;n15}
	Keen,~D.~A.; Goodwin,~A.~L. The crystallography of correlated disorder. \emph{Nature} \textbf{2015}, \emph{521}, 303--309\relax
	\mciteBstWouldAddEndPuncttrue
	\mciteSetBstMidEndSepPunct{\mcitedefaultmidpunct}
	{\mcitedefaultendpunct}{\mcitedefaultseppunct}\relax
	\EndOfBibitem
	\bibitem[Zhu \latin{et~al.}(2021)Zhu, Huang, Ren, Zhang, Ke, Jen, Zhang, Wang, and Liu]{zhu;advs21}
	Zhu,~H.; Huang,~Y.; Ren,~J.; Zhang,~B.; Ke,~Y.; Jen,~A. K.-Y.; Zhang,~Q.; Wang,~X.-L.; Liu,~Q. {Bridging Structural Inhomogeneity to Functionality: Pair Distribution Function Methods for Functional Materials Development}. \emph{Adv. Sci.} \textbf{2021}, \emph{8}, 2003534\relax
	\mciteBstWouldAddEndPuncttrue
	\mciteSetBstMidEndSepPunct{\mcitedefaultmidpunct}
	{\mcitedefaultendpunct}{\mcitedefaultseppunct}\relax
	\EndOfBibitem
	\bibitem[Zunger(2022)]{zunge;ncs22}
	Zunger,~A. {Bridging the gap between density functional theory and quantum materials}. \emph{Nat. Comput. Sci.} \textbf{2022}, \emph{2}, 529 -- 532\relax
	\mciteBstWouldAddEndPuncttrue
	\mciteSetBstMidEndSepPunct{\mcitedefaultmidpunct}
	{\mcitedefaultendpunct}{\mcitedefaultseppunct}\relax
	\EndOfBibitem
	\bibitem[Kimber \latin{et~al.}(2023)Kimber, Zhang, Liang, Guzm\'an-Verri, Littlewood, Cheng, Abernathy, Hudspeth, Luo, Kanatzidis, Chatterji, Ramirez-Cuesta, and Billinge]{kimbe;nm23}
	Kimber,~S. A.~J.; Zhang,~J.; Liang,~C.~H.; Guzm\'an-Verri,~G.~G.; Littlewood,~P.~B.; Cheng,~Y.; Abernathy,~D.~L.; Hudspeth,~J.~M.; Luo,~Z.-Z.; Kanatzidis,~M.~G.; Chatterji,~T.; Ramirez-Cuesta,~A.~J.; Billinge,~S. J.~L. {Dynamic crystallography reveals spontaneous anisotropy in cubic GeTe}. \emph{Nature Mater.} \textbf{2023}, \emph{22}, 311--315\relax
	\mciteBstWouldAddEndPuncttrue
	\mciteSetBstMidEndSepPunct{\mcitedefaultmidpunct}
	{\mcitedefaultendpunct}{\mcitedefaultseppunct}\relax
	\EndOfBibitem
	\bibitem[Tucker \latin{et~al.}(2005)Tucker, Goodwin, Dove, Keen, Wells, and Evans]{Tucker+etal2005}
	Tucker,~M.~G.; Goodwin,~A.~L.; Dove,~M.~T.; Keen,~D.~A.; Wells,~S.~A.; Evans,~J. S.~O. Negative Thermal Expansion in ZrW$_2$O$_8$ : Mechanisms, Rigid Unit Modes, and Neutron Total Scattering. \emph{{P}hys. {R}ev. {L}ett.} \textbf{2005}, \emph{95}, 255501\relax
	\mciteBstWouldAddEndPuncttrue
	\mciteSetBstMidEndSepPunct{\mcitedefaultmidpunct}
	{\mcitedefaultendpunct}{\mcitedefaultseppunct}\relax
	\EndOfBibitem
	\bibitem[Wang \latin{et~al.}(2017)Wang, Zhang, Chen, Shi, and Liu]{wang;stam17}
	Wang,~Y.; Zhang,~W.; Chen,~L.; Shi,~S.; Liu,~J. {Quantitative description on structure–property relationships of Li-ion battery materials for high-throughput computations}. \emph{Sci. Technol. Adv. Mater.} \textbf{2017}, \emph{18}, 134--146\relax
	\mciteBstWouldAddEndPuncttrue
	\mciteSetBstMidEndSepPunct{\mcitedefaultmidpunct}
	{\mcitedefaultendpunct}{\mcitedefaultseppunct}\relax
	\EndOfBibitem
	\bibitem[Liu \latin{et~al.}(2022)Liu, Susilo, Lee, Abeykoon, Tong, Hu, Stavitski, Attenkofer, Ke, Chen, and Petrovic]{liu;acsn22}
	Liu,~Y.; Susilo,~R.~A.; Lee,~Y.; Abeykoon,~A. M.~M.; Tong,~X.; Hu,~Z.; Stavitski,~E.; Attenkofer,~K.; Ke,~L.; Chen,~B.; Petrovic,~C. {Short-Range Crystalline Order-Tuned Conductivity in Cr$_2$Si$_2$Te$_6$ van der Waals Magnetic Crystals}. \emph{ACS Nano} \textbf{2022}, \emph{16}, 13134--13143\relax
	\mciteBstWouldAddEndPuncttrue
	\mciteSetBstMidEndSepPunct{\mcitedefaultmidpunct}
	{\mcitedefaultendpunct}{\mcitedefaultseppunct}\relax
	\EndOfBibitem
	\bibitem[Li \latin{et~al.}(2018)Li, Zhang, Damjanovic, Chen, and Shrout]{li;afm18}
	Li,~F.; Zhang,~S.; Damjanovic,~D.; Chen,~L.-Q.; Shrout,~T.~R. {Local Structural Heterogeneity and Electromechanical Responses of Ferroelectrics: Learning from Relaxor Ferroelectrics}. \emph{Adv. Funct. Mater.} \textbf{2018}, \emph{28}, 1801504\relax
	\mciteBstWouldAddEndPuncttrue
	\mciteSetBstMidEndSepPunct{\mcitedefaultmidpunct}
	{\mcitedefaultendpunct}{\mcitedefaultseppunct}\relax
	\EndOfBibitem
	\bibitem[Paddison \latin{et~al.}(2018)Paddison, Gutmann, Stewart, Tucker, Dove, Keen, and Goodwin]{paddi;prb18}
	Paddison,~J. A.~M.; Gutmann,~M.~J.; Stewart,~J.~R.; Tucker,~M.~G.; Dove,~M.~T.; Keen,~D.~A.; Goodwin,~A.~L. {Magnetic structure of paramagnetic MnO}. \emph{Phys. Rev. B} \textbf{2018}, \emph{97}, 014429\relax
	\mciteBstWouldAddEndPuncttrue
	\mciteSetBstMidEndSepPunct{\mcitedefaultmidpunct}
	{\mcitedefaultendpunct}{\mcitedefaultseppunct}\relax
	\EndOfBibitem
	\bibitem[Qureshi \latin{et~al.}(2022)Qureshi, Fischer, Riberolles, Hansen, Hatnean, and Petrenko]{Qureshi+etal2022}
	Qureshi,~N.; Fischer,~H.~E.; Riberolles,~S. X.~M.; Hansen,~T.~C.; Hatnean,~M.~C.; Petrenko,~O.~A. Magnetic short-range order in polycrystalline \ce{SrGd2O4} and \ce{SrNd2O4} studied by reverse Monte Carlo simulations and magnetic pair-distribution function analysis. \emph{{P}hys. {R}ev. {B}} \textbf{2022}, \emph{106}, 224426\relax
	\mciteBstWouldAddEndPuncttrue
	\mciteSetBstMidEndSepPunct{\mcitedefaultmidpunct}
	{\mcitedefaultendpunct}{\mcitedefaultseppunct}\relax
	\EndOfBibitem
	\bibitem[Fischer \latin{et~al.}(2002)Fischer, Kaul, and Kronm\"uller]{fisch;prb02}
	Fischer,~S.~F.; Kaul,~S.~N.; Kronm\"uller,~H. Critical magnetic properties of disordered polycrystalline ${\mathrm{Cr}}_{75}{\mathrm{Fe}}_{25}$ and ${\mathrm{Cr}}_{70}{\mathrm{Fe}}_{30}$ alloys. \emph{Phys. Rev. B} \textbf{2002}, \emph{65}, 064443\relax
	\mciteBstWouldAddEndPuncttrue
	\mciteSetBstMidEndSepPunct{\mcitedefaultmidpunct}
	{\mcitedefaultendpunct}{\mcitedefaultseppunct}\relax
	\EndOfBibitem
	\bibitem[Dietl and Ohno(2014)Dietl, and Ohno]{dietl;rmp14}
	Dietl,~T.; Ohno,~H. {Dilute ferromagnetic semiconductors: Physics and spintronic structures}. \emph{Rev. Mod. Phys.} \textbf{2014}, \emph{86}, 187--251\relax
	\mciteBstWouldAddEndPuncttrue
	\mciteSetBstMidEndSepPunct{\mcitedefaultmidpunct}
	{\mcitedefaultendpunct}{\mcitedefaultseppunct}\relax
	\EndOfBibitem
	\bibitem[Mydosh(2015)]{mydos;rpp15}
	Mydosh,~J.~A. {Spin glasses: redux: an updated experimental/materials survey}. \emph{Rep. Prog. Phys.} \textbf{2015}, \emph{78}, 052501\relax
	\mciteBstWouldAddEndPuncttrue
	\mciteSetBstMidEndSepPunct{\mcitedefaultmidpunct}
	{\mcitedefaultendpunct}{\mcitedefaultseppunct}\relax
	\EndOfBibitem
	\bibitem[Clark and Abdeldaim(2021)Clark, and Abdeldaim]{clark;armr21}
	Clark,~L.; Abdeldaim,~A.~H. {Quantum Spin Liquids from a Materials Perspective}. \emph{Annu. Rev. Mater. Res.} \textbf{2021}, \emph{51}, 495--519\relax
	\mciteBstWouldAddEndPuncttrue
	\mciteSetBstMidEndSepPunct{\mcitedefaultmidpunct}
	{\mcitedefaultendpunct}{\mcitedefaultseppunct}\relax
	\EndOfBibitem
	\bibitem[Zheng \latin{et~al.}(2019)Zheng, Lu, Polash, Rasoulianboroujeni, Liu, Manley, Deng, Sun, Chen, Hermann, Vashaee, Heremans, and Zhao]{zheng;sadv19}
	Zheng,~Y.; Lu,~T.; Polash,~M. M.~H.; Rasoulianboroujeni,~M.; Liu,~N.; Manley,~M.~E.; Deng,~Y.; Sun,~P.~J.; Chen,~X.~L.; Hermann,~R.~P.; Vashaee,~D.; Heremans,~J.~P.; Zhao,~H. {Paramagnon drag in high thermoelectric figure of merit Li-doped MnTe}. \emph{Sci. Adv.} \textbf{2019}, \emph{5}, eaat9461\relax
	\mciteBstWouldAddEndPuncttrue
	\mciteSetBstMidEndSepPunct{\mcitedefaultmidpunct}
	{\mcitedefaultendpunct}{\mcitedefaultseppunct}\relax
	\EndOfBibitem
	\bibitem[Miao \latin{et~al.}(2018)Miao, Hu, and Br\"uck]{miao;rm18}
	Miao,~X.-F.; Hu,~S.-Y.; Br\"uck,~E. {Overview of magnetoelastic coupling in (Mn,Fe)$_2$(P,Si)-type magnetocaloric materials}. \emph{Rare Met.} \textbf{2018}, \emph{37}, 723--733\relax
	\mciteBstWouldAddEndPuncttrue
	\mciteSetBstMidEndSepPunct{\mcitedefaultmidpunct}
	{\mcitedefaultendpunct}{\mcitedefaultseppunct}\relax
	\EndOfBibitem
	\bibitem[Walsh \latin{et~al.}(2021)Walsh, Asta, and Ritchie]{walsh;pnas21}
	Walsh,~F.; Asta,~M.; Ritchie,~R.~O. {Magnetically driven short-range order can explain anomalous measurements in CrCoNi}. \emph{Proc. Natl. Acad. Sci. USA} \textbf{2021}, \emph{118}, e2020540118\relax
	\mciteBstWouldAddEndPuncttrue
	\mciteSetBstMidEndSepPunct{\mcitedefaultmidpunct}
	{\mcitedefaultendpunct}{\mcitedefaultseppunct}\relax
	\EndOfBibitem
	\bibitem[van~der Minne \latin{et~al.}(2024)van~der Minne, Korol, Krakers, Verhage, Ros/'ario, Roskamp, Spiteri, Biz, Fianchini, Boukamp, Rijnders, Flipse, Gracia, Mul, Hilgenkamp, Green, Koster, and Baeumer]{minne;aprev24}
	van~der Minne,~E. \latin{et~al.}  {The effect of intrinsic magnetic order on electrochemical water splitting}. \emph{Appl. Phys. Rev.} \textbf{2024}, \emph{11}, 011420\relax
	\mciteBstWouldAddEndPuncttrue
	\mciteSetBstMidEndSepPunct{\mcitedefaultmidpunct}
	{\mcitedefaultendpunct}{\mcitedefaultseppunct}\relax
	\EndOfBibitem
	\bibitem[Jones \latin{et~al.}(2024)Jones, Suggs, Krivyakina, Phelan, Garlea, Chmaissem, and Frandsen]{jones;prb24}
	Jones,~B.; Suggs,~C.~Z.; Krivyakina,~E.; Phelan,~D.; Garlea,~V.~O.; Chmaissem,~O.; Frandsen,~B.~A. {Local atomic and magnetic structure of multiferroic (Sr, Ba)(Mn, Ti)O$_3$}. \emph{Phys. Rev. B} \textbf{2024}, \emph{109}, 024423\relax
	\mciteBstWouldAddEndPuncttrue
	\mciteSetBstMidEndSepPunct{\mcitedefaultmidpunct}
	{\mcitedefaultendpunct}{\mcitedefaultseppunct}\relax
	\EndOfBibitem
	\bibitem[Bean and Rodbell(1962)Bean, and Rodbell]{bean;pr62}
	Bean,~C.~P.; Rodbell,~D.~S. {Magnetic Disorder as a First-Order Phase Transformation}. \emph{Phys. Rev.} \textbf{1962}, \emph{126}, 104--115\relax
	\mciteBstWouldAddEndPuncttrue
	\mciteSetBstMidEndSepPunct{\mcitedefaultmidpunct}
	{\mcitedefaultendpunct}{\mcitedefaultseppunct}\relax
	\EndOfBibitem
	\bibitem[Norman(2011)]{norma;s11}
	Norman,~M.~R. {The Challenge of Unconventional Superconductivity}. \emph{Science} \textbf{2011}, \emph{332}, 196--200\relax
	\mciteBstWouldAddEndPuncttrue
	\mciteSetBstMidEndSepPunct{\mcitedefaultmidpunct}
	{\mcitedefaultendpunct}{\mcitedefaultseppunct}\relax
	\EndOfBibitem
	\bibitem[Frandsen \latin{et~al.}(2014)Frandsen, Yang, and Billinge]{Frandsen+etal2014}
	Frandsen,~B.~A.; Yang,~X.; Billinge,~S. J.~L. Magnetic pair distribution function analysis of local magnetic correlations. \emph{{A}cta {C}ryst.} \textbf{2014}, \emph{A70}, 3--11\relax
	\mciteBstWouldAddEndPuncttrue
	\mciteSetBstMidEndSepPunct{\mcitedefaultmidpunct}
	{\mcitedefaultendpunct}{\mcitedefaultseppunct}\relax
	\EndOfBibitem
	\bibitem[Cava \latin{et~al.}(2021)Cava, de~Leon, and Xie]{cava;cr21}
	Cava,~R.; de~Leon,~N.; Xie,~W. {Introduction: Quantum Materials}. \emph{Chem. Rev.} \textbf{2021}, \emph{121}, 2777--2779\relax
	\mciteBstWouldAddEndPuncttrue
	\mciteSetBstMidEndSepPunct{\mcitedefaultmidpunct}
	{\mcitedefaultendpunct}{\mcitedefaultseppunct}\relax
	\EndOfBibitem
	\bibitem[Keimer \latin{et~al.}(2015)Keimer, Kivelson, Norman, Uchida, and Zaanen]{keime;n15}
	Keimer,~B.; Kivelson,~S.~A.; Norman,~M.~R.; Uchida,~S.; Zaanen,~J. From quantum matter to high-temperature superconductivity in copper oxides. \emph{Nature} \textbf{2015}, \emph{518}, 179--186\relax
	\mciteBstWouldAddEndPuncttrue
	\mciteSetBstMidEndSepPunct{\mcitedefaultmidpunct}
	{\mcitedefaultendpunct}{\mcitedefaultseppunct}\relax
	\EndOfBibitem
	\bibitem[Narang \latin{et~al.}(2021)Narang, Garcia, and Felser]{naran;nm21}
	Narang,~P.; Garcia,~C. A.~C.; Felser,~C. {The topology of electronic band structures}. \emph{Nat. Mater.} \textbf{2021}, \emph{20}, 293--300\relax
	\mciteBstWouldAddEndPuncttrue
	\mciteSetBstMidEndSepPunct{\mcitedefaultmidpunct}
	{\mcitedefaultendpunct}{\mcitedefaultseppunct}\relax
	\EndOfBibitem
	\bibitem[Ramirez(1994)]{ramir;aroms94}
	Ramirez,~A.~P. {Strongly Geometrically Frustrated Magnets}. \emph{Annu. Rev. Mater. Sci.} \textbf{1994}, \emph{24}, 453--480\relax
	\mciteBstWouldAddEndPuncttrue
	\mciteSetBstMidEndSepPunct{\mcitedefaultmidpunct}
	{\mcitedefaultendpunct}{\mcitedefaultseppunct}\relax
	\EndOfBibitem
	\bibitem[Balents(2010)]{balen;n10}
	Balents,~L. Spin liquids in frustrated magnets. \emph{Nature} \textbf{2010}, \emph{464}, 199--208\relax
	\mciteBstWouldAddEndPuncttrue
	\mciteSetBstMidEndSepPunct{\mcitedefaultmidpunct}
	{\mcitedefaultendpunct}{\mcitedefaultseppunct}\relax
	\EndOfBibitem
	\bibitem[Castelnovo \latin{et~al.}(2012)Castelnovo, Moessner, and Sondhi]{caste;arocmp12}
	Castelnovo,~C.; Moessner,~R.; Sondhi,~S.~L. {Spin Ice, Fractionalization, and Topological Order}. \emph{Annu. Rev. Condens. Matter Phys.} \textbf{2012}, \emph{3}, 35--55\relax
	\mciteBstWouldAddEndPuncttrue
	\mciteSetBstMidEndSepPunct{\mcitedefaultmidpunct}
	{\mcitedefaultendpunct}{\mcitedefaultseppunct}\relax
	\EndOfBibitem
	\bibitem[Savary and Balents(2017)Savary, and Balents]{savar;rpp17}
	Savary,~L.; Balents,~L. Quantum spin liquids: a review. \emph{Rep. Prog. Phys.} \textbf{2017}, \emph{80}, 016502\relax
	\mciteBstWouldAddEndPuncttrue
	\mciteSetBstMidEndSepPunct{\mcitedefaultmidpunct}
	{\mcitedefaultendpunct}{\mcitedefaultseppunct}\relax
	\EndOfBibitem
	\bibitem[Broholm \latin{et~al.}(2020)Broholm, Cava, Kivelson, Nocera, Norman, and Senthil]{broho;s20}
	Broholm,~C.; Cava,~R.~J.; Kivelson,~S.~A.; Nocera,~D.~G.; Norman,~M.~R.; Senthil,~T. Quantum spin liquids. \emph{Science} \textbf{2020}, \emph{367}, eaay0668\relax
	\mciteBstWouldAddEndPuncttrue
	\mciteSetBstMidEndSepPunct{\mcitedefaultmidpunct}
	{\mcitedefaultendpunct}{\mcitedefaultseppunct}\relax
	\EndOfBibitem
	\bibitem[Anderson(1973)]{ander;mrb73}
	Anderson,~P.~W. {Resonating valence bonds: A new kind of insulator?} \emph{Mater. Res. Bull.} \textbf{1973}, \emph{8}, 153--160\relax
	\mciteBstWouldAddEndPuncttrue
	\mciteSetBstMidEndSepPunct{\mcitedefaultmidpunct}
	{\mcitedefaultendpunct}{\mcitedefaultseppunct}\relax
	\EndOfBibitem
	\bibitem[Knolle and Moessner(2019)Knolle, and Moessner]{knoll;arocmp19}
	Knolle,~J.; Moessner,~R. {A Field Guide to Spin Liquids}. \emph{Annu. Rev. Condens. Matter Phys.} \textbf{2019}, \emph{10}, 451--472\relax
	\mciteBstWouldAddEndPuncttrue
	\mciteSetBstMidEndSepPunct{\mcitedefaultmidpunct}
	{\mcitedefaultendpunct}{\mcitedefaultseppunct}\relax
	\EndOfBibitem
	\bibitem[Mustonen \latin{et~al.}(2024)Mustonen, Fogh, Paddison, Mangin-Thro, Hansen, Playford, Diaz-Lopez, Babkevich, Vasala, Karppinen, Cussen, R{\o}nnow, and Walker]{musto;cm24}
	Mustonen,~O. H.~J.; Fogh,~E.; Paddison,~J. A.~M.; Mangin-Thro,~L.; Hansen,~T.; Playford,~H.~Y.; Diaz-Lopez,~M.; Babkevich,~P.; Vasala,~S.; Karppinen,~M.; Cussen,~E.~J.; R{\o}nnow,~H.~M.; Walker,~H.~C. {Structure, Spin Correlations, and Magnetism of the $S = 1/2$ Square-Lattice Antiferromagnet Sr$_2$CuTe$_{1-x}$W$_x$O$_6$ ($0 \le x \le 1$)}. \emph{Chem. Mater.} \textbf{2024}, \emph{36}, 501--513\relax
	\mciteBstWouldAddEndPuncttrue
	\mciteSetBstMidEndSepPunct{\mcitedefaultmidpunct}
	{\mcitedefaultendpunct}{\mcitedefaultseppunct}\relax
	\EndOfBibitem
	\bibitem[Nuttall \latin{et~al.}(2023)Nuttall, Suggs, Fischer, Bordelon, Wilson, and Frandsen]{Nuttall+etal2023}
	Nuttall,~K.~M.; Suggs,~C.~Z.; Fischer,~H.~E.; Bordelon,~M.~M.; Wilson,~S.~D.; Frandsen,~B.~A. Quantitative investigation of the short-range magnetic correlations in the candidate quantum spin liquid \ce{NaYbO2}. \emph{{P}hys. {R}ev. {B}} \textbf{2023}, \emph{108}, L140411\relax
	\mciteBstWouldAddEndPuncttrue
	\mciteSetBstMidEndSepPunct{\mcitedefaultmidpunct}
	{\mcitedefaultendpunct}{\mcitedefaultseppunct}\relax
	\EndOfBibitem
	\bibitem[Scheie \latin{et~al.}(2024)Scheie, Ghioldi, Xing, Paddison, Sherman, Dupont, Sanjeewa, Lee, Woods, Abernathy, Pajerowski, Williams, Zhang, Manuel, Trumper, Pemmaraju, Sefat, Parker, Devereaux, Movshovich, Moore, Batista, and Tennant]{schei;np24}
	Scheie,~A.~O. \latin{et~al.}  {Proximate spin liquid and fractionalization in the triangular antiferromagnet KYbSe$_2$}. \emph{Nature Phys.} \textbf{2024}, \emph{20}, 74--81\relax
	\mciteBstWouldAddEndPuncttrue
	\mciteSetBstMidEndSepPunct{\mcitedefaultmidpunct}
	{\mcitedefaultendpunct}{\mcitedefaultseppunct}\relax
	\EndOfBibitem
	\bibitem[Burch \latin{et~al.}(2018)Burch, Mandrus, and Park]{burch;n18}
	Burch,~K.~S.; Mandrus,~D.; Park,~J.~G. {Magnetism in two-dimensional van der Waals materials}. \emph{Nature} \textbf{2018}, \emph{563}, 47--52\relax
	\mciteBstWouldAddEndPuncttrue
	\mciteSetBstMidEndSepPunct{\mcitedefaultmidpunct}
	{\mcitedefaultendpunct}{\mcitedefaultseppunct}\relax
	\EndOfBibitem
	\bibitem[Wang \latin{et~al.}(2020)Wang, Huang, Cheung, Chen, Tan, Huang, Zhao, Zhao, Wu, Feng, Wu, and Chang]{wang;adp20}
	Wang,~M.~C.; Huang,~C.~C.; Cheung,~C.~H.; Chen,~C.~Y.; Tan,~S.~G.; Huang,~T.~W.; Zhao,~Y.; Zhao,~Y.; Wu,~G.; Feng,~Y.~P.; Wu,~H.~C.; Chang,~C.~R. {Prospects and Opportunities of 2D van der Waals Magnetic Systems}. \emph{Ann. Phys.} \textbf{2020}, \emph{532}, 1900452\relax
	\mciteBstWouldAddEndPuncttrue
	\mciteSetBstMidEndSepPunct{\mcitedefaultmidpunct}
	{\mcitedefaultendpunct}{\mcitedefaultseppunct}\relax
	\EndOfBibitem
	\bibitem[Yang \latin{et~al.}(2021)Yang, Zhang, and Jiang]{yang;advs21}
	Yang,~S.; Zhang,~T.; Jiang,~C. {van der Waals Magnets: Material Family, Detection and Modulation of Magnetism, and Perspective in Spintronics}. \emph{Adv. Sci.} \textbf{2021}, \emph{8}, 2002488\relax
	\mciteBstWouldAddEndPuncttrue
	\mciteSetBstMidEndSepPunct{\mcitedefaultmidpunct}
	{\mcitedefaultendpunct}{\mcitedefaultseppunct}\relax
	\EndOfBibitem
	\bibitem[Wang \latin{et~al.}(2021)Wang, Bedoya-Pinto, Blei, Dismukes, Hamo, Jenkins, Koperski, Liu, Sun, Telford, Kim, Augustin, Vool, Yin, Li, Falin, Dean, Casanova, Evans, Chshiev, Mishchenko, Petrovic, He, Zhao, Tsen, Gerardot, Brotons-Gisbert, Guguchia, Roy, Tongay, Wang, Hasan, Wrachtrup, Yacoby, Fert, Parkin, Novoselov, Dai, Balicas, and Santos]{wang;acsn21}
	Wang,~Q.~H. \latin{et~al.}  {The Magnetic Genome of Two-Dimensional van der Waals Materials}. \emph{ACS Nano} \textbf{2021}, \emph{16}, 6960--7079\relax
	\mciteBstWouldAddEndPuncttrue
	\mciteSetBstMidEndSepPunct{\mcitedefaultmidpunct}
	{\mcitedefaultendpunct}{\mcitedefaultseppunct}\relax
	\EndOfBibitem
	\bibitem[Ren and Xiang(2023)Ren, and Xiang]{ren;nanom23}
	Ren,~H.; Xiang,~G. {Strain Engineering of Intrinsic Ferromagnetism in 2D van der Waals Materials}. \emph{Nanomaterials} \textbf{2023}, \emph{13}, 2378\relax
	\mciteBstWouldAddEndPuncttrue
	\mciteSetBstMidEndSepPunct{\mcitedefaultmidpunct}
	{\mcitedefaultendpunct}{\mcitedefaultseppunct}\relax
	\EndOfBibitem
	\bibitem[Mermin and Wagner(1966)Mermin, and Wagner]{mermi;prl66}
	Mermin,~N.~D.; Wagner,~H. {Absence of Ferromagnetism or Antiferromagnetism in One- or Two-Dimensional Isotropic Heisenberg Models}. \emph{Phys. Rev. Lett.} \textbf{1966}, \emph{17}, 1133--1136\relax
	\mciteBstWouldAddEndPuncttrue
	\mciteSetBstMidEndSepPunct{\mcitedefaultmidpunct}
	{\mcitedefaultendpunct}{\mcitedefaultseppunct}\relax
	\EndOfBibitem
	\bibitem[Yang \latin{et~al.}(2023)Yang, Wu, Li, Ran, Wang, Zhu, Gong, Liu, Wang, Zhang, Mi, Wang, Chai, Su, Wang, He, Yang, and Zhou]{yang;afm23}
	Yang,~K. \latin{et~al.}  {Spin-Phonon Scattering-Induced Low Thermal Conductivity in a van der Waals Layered Ferromagnet Cr$_2$Si$_2$Te$_6$}. \emph{Adv. Funct. Mater.} \textbf{2023}, \emph{33}, 2302191\relax
	\mciteBstWouldAddEndPuncttrue
	\mciteSetBstMidEndSepPunct{\mcitedefaultmidpunct}
	{\mcitedefaultendpunct}{\mcitedefaultseppunct}\relax
	\EndOfBibitem
	\bibitem[Williams \latin{et~al.}(2015)Williams, Aczel, Lumsden, Nagler, Stone, Yan, and Mandrus]{willi;prb15}
	Williams,~T.~J.; Aczel,~A.~A.; Lumsden,~M.~D.; Nagler,~S.~E.; Stone,~M.~B.; Yan,~J.-Q.; Mandrus,~D. {Magnetic correlations in the quasi-two-dimensional semiconducting ferromagnet CrSiTe$_3$}. \emph{Phys. Rev. B} \textbf{2015}, \emph{92}, 144404\relax
	\mciteBstWouldAddEndPuncttrue
	\mciteSetBstMidEndSepPunct{\mcitedefaultmidpunct}
	{\mcitedefaultendpunct}{\mcitedefaultseppunct}\relax
	\EndOfBibitem
	\bibitem[Keyes(1959)]{KeyesPR1959}
	Keyes,~R.~W. {High-Temperature Thermal Conductivity of Insulating Crystals: Relationship to the Melting Point}. \emph{Phys. Rev.} \textbf{1959}, \emph{115}, 564--567\relax
	\mciteBstWouldAddEndPuncttrue
	\mciteSetBstMidEndSepPunct{\mcitedefaultmidpunct}
	{\mcitedefaultendpunct}{\mcitedefaultseppunct}\relax
	\EndOfBibitem
	\bibitem[Kitanovski(2020)]{kitan;aem20}
	Kitanovski,~A. {Energy Applications of Magnetocaloric Materials}. \emph{Adv. Energy Mater.} \textbf{2020}, \emph{10}, 1903741\relax
	\mciteBstWouldAddEndPuncttrue
	\mciteSetBstMidEndSepPunct{\mcitedefaultmidpunct}
	{\mcitedefaultendpunct}{\mcitedefaultseppunct}\relax
	\EndOfBibitem
	\bibitem[Fischer(1988)]{Magnetic_Cooling_in_LowT_Book_1988}
	Fischer,~H.~E. In \emph{Experimental Techniques In Condensed Matter Physics At Low Temperatures}, 1st ed.; Richardson,~R.~C., Smith,~E.~N., Eds.; Addison-Wesley/Frontiers in Physics Series vol.~67: Redwood City, California, 1988; pp 76--96\relax
	\mciteBstWouldAddEndPuncttrue
	\mciteSetBstMidEndSepPunct{\mcitedefaultmidpunct}
	{\mcitedefaultendpunct}{\mcitedefaultseppunct}\relax
	\EndOfBibitem
	\bibitem[Franco \latin{et~al.}(2012)Franco, Bl\'azquez, Ingale, and Conde]{franc;armr12}
	Franco,~V.; Bl\'azquez,~J.; Ingale,~B.; Conde,~A. {The Magnetocaloric Effect and Magnetic Refrigeration Near Room Temperature: Materials and Models}. \emph{Annu. Rev. Mater. Res.} \textbf{2012}, \emph{42}, 305--342\relax
	\mciteBstWouldAddEndPuncttrue
	\mciteSetBstMidEndSepPunct{\mcitedefaultmidpunct}
	{\mcitedefaultendpunct}{\mcitedefaultseppunct}\relax
	\EndOfBibitem
	\bibitem[Miao \latin{et~al.}(2016)Miao, Caron, Cedervall, Gubbens, Dalmas~de R\'eotier, Yaouanc, Qian, Wildes, Luetkens, Amato, van Dijk, and Br\"uck]{miao;prb16}
	Miao,~X.~F.; Caron,~L.; Cedervall,~J.; Gubbens,~P. C.~M.; Dalmas~de R\'eotier,~P.; Yaouanc,~A.; Qian,~F.; Wildes,~A.~R.; Luetkens,~H.; Amato,~A.; van Dijk,~N.~H.; Br\"uck,~E. {Short-range magnetic correlations and spin dynamics in the paramagnetic regime of (Mn,Fe)$_2$(P,Si)}. \emph{Phys. Rev. B} \textbf{2016}, \emph{94}, 014426\relax
	\mciteBstWouldAddEndPuncttrue
	\mciteSetBstMidEndSepPunct{\mcitedefaultmidpunct}
	{\mcitedefaultendpunct}{\mcitedefaultseppunct}\relax
	\EndOfBibitem
	\bibitem[Boeije \latin{et~al.}(2017)Boeije, Maschek, Miao, Thang, van Dijk, and Br\"uck]{boeij;jpd17}
	Boeije,~M. F.~J.; Maschek,~M.; Miao,~X.~F.; Thang,~N.~V.; van Dijk,~N.~H.; Br\"uck,~E. {Mixed magnetism in magnetocaloric materials with first-order and second-order magnetoelastic transitions}. \emph{J. Phys. D: Appl. Phys.} \textbf{2017}, \emph{50}, 174002\relax
	\mciteBstWouldAddEndPuncttrue
	\mciteSetBstMidEndSepPunct{\mcitedefaultmidpunct}
	{\mcitedefaultendpunct}{\mcitedefaultseppunct}\relax
	\EndOfBibitem
	\bibitem[Pakhira \latin{et~al.}(2017)Pakhira, Mazumdar, Ranganathan, and Avdeev]{pakhi;sr17}
	Pakhira,~S.; Mazumdar,~C.; Ranganathan,~R.; Avdeev,~M. Magnetic frustration induced large magnetocaloric effect in the absence of long range magnetic order. \emph{Sci. Rep.} \textbf{2017}, \emph{7}, 7367\relax
	\mciteBstWouldAddEndPuncttrue
	\mciteSetBstMidEndSepPunct{\mcitedefaultmidpunct}
	{\mcitedefaultendpunct}{\mcitedefaultseppunct}\relax
	\EndOfBibitem
	\bibitem[Muniraju \latin{et~al.}(2020)Muniraju, Baral, Tian, Li, Poudel, Gofryk, Bari\v{s}i\'{c}, Kiefer, Ross, and Nair]{munir;ic20}
	Muniraju,~N. K.~C.; Baral,~R.; Tian,~Y.; Li,~R.; Poudel,~N.; Gofryk,~K.; Bari\v{s}i\'{c},~N.; Kiefer,~B.; Ross,~J. H.~J.; Nair,~H.~S. {Magnetocaloric Effect in a Frustrated Gd-Garnet with No Long-Range Magnetic Order}. \emph{Inorg. Chem.} \textbf{2020}, \emph{59}, 15144--15153\relax
	\mciteBstWouldAddEndPuncttrue
	\mciteSetBstMidEndSepPunct{\mcitedefaultmidpunct}
	{\mcitedefaultendpunct}{\mcitedefaultseppunct}\relax
	\EndOfBibitem
	\bibitem[Shi \latin{et~al.}(2020)Shi, Zou, and Chen]{shi;cr20}
	Shi,~X.-L.; Zou,~J.; Chen,~Z.-G. {Advanced Thermoelectric Design: From Materials and Structures to Devices}. \emph{Chem. Rev.} \textbf{2020}, \emph{120}, 7399--7515\relax
	\mciteBstWouldAddEndPuncttrue
	\mciteSetBstMidEndSepPunct{\mcitedefaultmidpunct}
	{\mcitedefaultendpunct}{\mcitedefaultseppunct}\relax
	\EndOfBibitem
	\bibitem[Yang \latin{et~al.}(2023)Yang, Sang, Zhang, Ye, Hamilton, Fuhrer, and Wang]{yang;nrp23}
	Yang,~G.; Sang,~L.; Zhang,~C.; Ye,~N.; Hamilton,~A.; Fuhrer,~M.~S.; Wang,~X. The role of spin in thermoelectricity. \emph{Nat. Rev. Phys.} \textbf{2023}, \emph{5}, 466--482\relax
	\mciteBstWouldAddEndPuncttrue
	\mciteSetBstMidEndSepPunct{\mcitedefaultmidpunct}
	{\mcitedefaultendpunct}{\mcitedefaultseppunct}\relax
	\EndOfBibitem
	\bibitem[Liu \latin{et~al.}(2023)Liu, Chen, Fu, and Zhu]{liu;advpr23}
	Liu,~S.; Chen,~M.; Fu,~C.; Zhu,~T. {The Interplay of Magnetism and Thermoelectricity: A Review}. \emph{Adv. Phys. Res.} \textbf{2023}, \emph{2}, 2300015\relax
	\mciteBstWouldAddEndPuncttrue
	\mciteSetBstMidEndSepPunct{\mcitedefaultmidpunct}
	{\mcitedefaultendpunct}{\mcitedefaultseppunct}\relax
	\EndOfBibitem
	\bibitem[B.~M. and Guin(2023)B.~M., and Guin]{kumar;mcf23}
	B.~M.,~A.~K.; Guin,~S.~N. Topological quantum magnets for transverse thermoelectric energy conversion. \emph{Mater. Chem. Front.} \textbf{2023}, \emph{7}, 4202--4214\relax
	\mciteBstWouldAddEndPuncttrue
	\mciteSetBstMidEndSepPunct{\mcitedefaultmidpunct}
	{\mcitedefaultendpunct}{\mcitedefaultseppunct}\relax
	\EndOfBibitem
	\bibitem[Koshibae and Maekawa(2001)Koshibae, and Maekawa]{koshi;prl01}
	Koshibae,~W.; Maekawa,~S. {Effects of Spin and Orbital Degeneracy on the Thermopower of Strongly Correlated Systems}. \emph{Phys. Rev. Lett.} \textbf{2001}, \emph{87}, 236603\relax
	\mciteBstWouldAddEndPuncttrue
	\mciteSetBstMidEndSepPunct{\mcitedefaultmidpunct}
	{\mcitedefaultendpunct}{\mcitedefaultseppunct}\relax
	\EndOfBibitem
	\bibitem[Polash \latin{et~al.}(2021)Polash, Moseley, Zhang, Hermann, and Vashaee]{polas;crps21}
	Polash,~M. M.~H.; Moseley,~D.; Zhang,~J.; Hermann,~R.~P.; Vashaee,~D. {Understanding and design of spin-driven thermoelectrics}. \emph{Cell Rep. Phys. Sci.} \textbf{2021}, \emph{2}, 100614\relax
	\mciteBstWouldAddEndPuncttrue
	\mciteSetBstMidEndSepPunct{\mcitedefaultmidpunct}
	{\mcitedefaultendpunct}{\mcitedefaultseppunct}\relax
	\EndOfBibitem
	\bibitem[Wang \latin{et~al.}(2023)Wang, Jiang, Zhou, Wang, Wang, Chai, He, Han, Ying, Lu, Pan, Wang, Zhou, and Chen]{wang;prb23}
	Wang,~H.; Jiang,~L.; Zhou,~Z.; Wang,~R.; Wang,~A.; Chai,~Y.; He,~M.; Han,~G.; Ying,~J.; Lu,~X.; Pan,~Y.; Wang,~G.; Zhou,~X.; Chen,~X. {Magnetic frustration driven high thermoelectric performance in the kagome antiferromagnet YMn$_6$Sn$_6$}. \emph{Phys. Rev. B} \textbf{2023}, \emph{108}, 155135\relax
	\mciteBstWouldAddEndPuncttrue
	\mciteSetBstMidEndSepPunct{\mcitedefaultmidpunct}
	{\mcitedefaultendpunct}{\mcitedefaultseppunct}\relax
	\EndOfBibitem
	\bibitem[Baral \latin{et~al.}(2022)Baral, Christensen, Hamilton, Ye, Chesnel, Sparks, Ward, Yan, McGuire, Manley, Staunton, Hermann, and Frandsen]{baral;matter22}
	Baral,~R.; Christensen,~J.; Hamilton,~P.; Ye,~F.; Chesnel,~K.; Sparks,~T.~D.; Ward,~R.; Yan,~J.; McGuire,~M.~A.; Manley,~M.~E.; Staunton,~J.~B.; Hermann,~R.~P.; Frandsen,~B.~A. Real-space visualization of short-range antiferromagnetic correlations in a magnetically enhanced thermoelectric. \emph{Matter} \textbf{2022}, \emph{5}, 1853--1864\relax
	\mciteBstWouldAddEndPuncttrue
	\mciteSetBstMidEndSepPunct{\mcitedefaultmidpunct}
	{\mcitedefaultendpunct}{\mcitedefaultseppunct}\relax
	\EndOfBibitem
	\bibitem[Billinge \latin{et~al.}(2023)Billinge, Skjaervoe, Terban, Tao, Yang, Rakita, and Frandsen]{billi;chapter;cic23}
	Billinge,~S.~J.; Skjaervoe,~S.~H.; Terban,~M.~W.; Tao,~S.; Yang,~L.; Rakita,~Y.; Frandsen,~B.~A. In \emph{{Comprehensive Inorganic Chemistry III (Third Edition)}}, third edition ed.; Reedijk,~J., Poeppelmeier,~K.~R., Eds.; Elsevier: Oxford, 2023; pp 222--247\relax
	\mciteBstWouldAddEndPuncttrue
	\mciteSetBstMidEndSepPunct{\mcitedefaultmidpunct}
	{\mcitedefaultendpunct}{\mcitedefaultseppunct}\relax
	\EndOfBibitem
	\bibitem[Egami and Billinge(2012)Egami, and Billinge]{Egami+Billinge_book_2012}
	Egami,~T.; Billinge,~S. J.~L. \emph{Underneath the Bragg Peaks: Structural Analysis of Complex Materials}, 2nd ed.; Elsevier/Pergamon Materials Series vol.~16: Oxford, United Kingdom, 2012\relax
	\mciteBstWouldAddEndPuncttrue
	\mciteSetBstMidEndSepPunct{\mcitedefaultmidpunct}
	{\mcitedefaultendpunct}{\mcitedefaultseppunct}\relax
	\EndOfBibitem
	\bibitem[Fischer \latin{et~al.}(2006)Fischer, Barnes, and Salmon]{Fischer+etal_Review_2006}
	Fischer,~H.~E.; Barnes,~A.~C.; Salmon,~P.~S. Neutron and x-ray diffraction studies of liquids and glasses. \emph{Rep. Prog. Phys.} \textbf{2006}, \emph{69}, 233 -- 299\relax
	\mciteBstWouldAddEndPuncttrue
	\mciteSetBstMidEndSepPunct{\mcitedefaultmidpunct}
	{\mcitedefaultendpunct}{\mcitedefaultseppunct}\relax
	\EndOfBibitem
	\bibitem[Terban and Billinge(2022)Terban, and Billinge]{Terban+Billinge_2022}
	Terban,~M.~W.; Billinge,~S. J.~L. Structural Analysis of Molecular Materials Using the Pair Distribution Function. \emph{Chem. Rev.} \textbf{2022}, \emph{122}, 1208 -- 1272\relax
	\mciteBstWouldAddEndPuncttrue
	\mciteSetBstMidEndSepPunct{\mcitedefaultmidpunct}
	{\mcitedefaultendpunct}{\mcitedefaultseppunct}\relax
	\EndOfBibitem
	\bibitem[Keen(2020)]{Keen2020}
	Keen,~D.~A. Total scattering and the pair distribution function in crystallography. \emph{{C}rystallography {R}eviews} \textbf{2020}, \emph{26}, 143--201\relax
	\mciteBstWouldAddEndPuncttrue
	\mciteSetBstMidEndSepPunct{\mcitedefaultmidpunct}
	{\mcitedefaultendpunct}{\mcitedefaultseppunct}\relax
	\EndOfBibitem
	\bibitem[Peterson and Keen(2021)Peterson, and Keen]{Peterson+Keen_2021}
	Peterson,~P.~F.; Keen,~D.~A. Illustrated formalisms for total scattering data: a guide for new practitioners. Corrigendum and addendum. \emph{J. {A}ppl. {C}ryst.} \textbf{2021}, \emph{54}, 1542 -- 1545\relax
	\mciteBstWouldAddEndPuncttrue
	\mciteSetBstMidEndSepPunct{\mcitedefaultmidpunct}
	{\mcitedefaultendpunct}{\mcitedefaultseppunct}\relax
	\EndOfBibitem
	\bibitem[Farrow and Billinge(2009)Farrow, and Billinge]{Farrow+Billinge2009}
	Farrow,~C.~L.; Billinge,~S. J.~L. Relationship between the atomic pair distribution function and small-angle scattering: implications for modeling of nanoparticles. \emph{{A}cta {C}ryst.} \textbf{2009}, \emph{A65}, 232--239\relax
	\mciteBstWouldAddEndPuncttrue
	\mciteSetBstMidEndSepPunct{\mcitedefaultmidpunct}
	{\mcitedefaultendpunct}{\mcitedefaultseppunct}\relax
	\EndOfBibitem
	\bibitem[G\"ahler \latin{et~al.}(1998)G\"ahler, Felber, Mezei, and Golub]{Gaehler+etal1998}
	G\"ahler,~R.; Felber,~J.; Mezei,~F.; Golub,~R. Space-time approach to scattering from many-body systems. \emph{{P}hys. {R}ev. {B}} \textbf{1998}, \emph{58}, 280\relax
	\mciteBstWouldAddEndPuncttrue
	\mciteSetBstMidEndSepPunct{\mcitedefaultmidpunct}
	{\mcitedefaultendpunct}{\mcitedefaultseppunct}\relax
	\EndOfBibitem
	\bibitem[Desgranges \latin{et~al.}(2023)Desgranges, Baldinozzi, Fischer, and Lander]{Desgranges+etal_2023}
	Desgranges,~L.; Baldinozzi,~G.; Fischer,~H.~E.; Lander,~G.~H. Temperature-dependent anisotropy in the bond lengths of \ce{UO2} as a result of phonon-induced atomic correlations. \emph{J. Phys.: Condens. Matter} \textbf{2023}, \emph{35}, 10LT01\relax
	\mciteBstWouldAddEndPuncttrue
	\mciteSetBstMidEndSepPunct{\mcitedefaultmidpunct}
	{\mcitedefaultendpunct}{\mcitedefaultseppunct}\relax
	\EndOfBibitem
	\bibitem[McQueeney(1998)]{McQueeney1998}
	McQueeney,~R.~J. {Dynamic radial distribution function from inelastic neutron scattering}. \emph{Phys. Rev. B} \textbf{1998}, \emph{57}, 10560--10568\relax
	\mciteBstWouldAddEndPuncttrue
	\mciteSetBstMidEndSepPunct{\mcitedefaultmidpunct}
	{\mcitedefaultendpunct}{\mcitedefaultseppunct}\relax
	\EndOfBibitem
	\bibitem[Fry-Petit \latin{et~al.}(2015)Fry-Petit, Rebola, Mourigal, Valentine, Drichko, Sheckelton, Fennie, and McQueen]{Fry-Petit2015}
	Fry-Petit,~A.~M.; Rebola,~A.~F.; Mourigal,~M.; Valentine,~M.; Drichko,~N.; Sheckelton,~J.~P.; Fennie,~C.~J.; McQueen,~T.~M. {Direct assignment of molecular vibrations via normal mode analysis of the neutron dynamic pair distribution function technique}. \emph{J. Chem. Phys.} \textbf{2015}, \emph{143}, 124201\relax
	\mciteBstWouldAddEndPuncttrue
	\mciteSetBstMidEndSepPunct{\mcitedefaultmidpunct}
	{\mcitedefaultendpunct}{\mcitedefaultseppunct}\relax
	\EndOfBibitem
	\bibitem[Wedgwood and Wright(1976)Wedgwood, and Wright]{Wedgwood+Wright1976}
	Wedgwood,~F.~A.; Wright,~A.~C. {Short range antiferromagnetic ordering in vitreous \ce{Fe2O3-P2O5}}. \emph{J. Non-Cryst. Solids} \textbf{1976}, \emph{21}, 95--105\relax
	\mciteBstWouldAddEndPuncttrue
	\mciteSetBstMidEndSepPunct{\mcitedefaultmidpunct}
	{\mcitedefaultendpunct}{\mcitedefaultseppunct}\relax
	\EndOfBibitem
	\bibitem[Wright(1980)]{Wright1980}
	Wright,~A.~C. {Neutron magnetic scattering studies of amorphous solids}. \emph{J. Non-Cryst. Solids} \textbf{1980}, \emph{40}, 325--346\relax
	\mciteBstWouldAddEndPuncttrue
	\mciteSetBstMidEndSepPunct{\mcitedefaultmidpunct}
	{\mcitedefaultendpunct}{\mcitedefaultseppunct}\relax
	\EndOfBibitem
	\bibitem[Keen and McGreevy(1991)Keen, and McGreevy]{Keen+McGreevy1991}
	Keen,~D.~A.; McGreevy,~R.~L. Determination of disordered magnetic structures by RMC modelling of neutron diffraction data. \emph{J. Phys.: Condens. Matter} \textbf{1991}, \emph{3}, 7383--7394\relax
	\mciteBstWouldAddEndPuncttrue
	\mciteSetBstMidEndSepPunct{\mcitedefaultmidpunct}
	{\mcitedefaultendpunct}{\mcitedefaultseppunct}\relax
	\EndOfBibitem
	\bibitem[Paddison \latin{et~al.}(2013)Paddison, Stewart, and Goodwin]{Paddison+etal_Spinvert_2013}
	Paddison,~J. A.~M.; Stewart,~J.~R.; Goodwin,~A.~L. Spinvert: a program for refinement of paramagnetic diffuse scattering data. \emph{J. Phys.: Condens. Matter} \textbf{2013}, \emph{25}, 454220\relax
	\mciteBstWouldAddEndPuncttrue
	\mciteSetBstMidEndSepPunct{\mcitedefaultmidpunct}
	{\mcitedefaultendpunct}{\mcitedefaultseppunct}\relax
	\EndOfBibitem
	\bibitem[Kodama \latin{et~al.}(2017)Kodama, Ikeda, Shamoto, and Otomo]{Kodama+etal2017}
	Kodama,~K.; Ikeda,~K.; Shamoto,~S.-i.; Otomo,~T. Alternative Equation on Magnetic Pair Distribution Function for Quantitative Analysis. \emph{J. Phys. Soc. Jpn.} \textbf{2017}, \emph{86}, 124708\relax
	\mciteBstWouldAddEndPuncttrue
	\mciteSetBstMidEndSepPunct{\mcitedefaultmidpunct}
	{\mcitedefaultendpunct}{\mcitedefaultseppunct}\relax
	\EndOfBibitem
	\bibitem[Qureshi(2019)]{Qureshi_Mag2Pol_2019}
	Qureshi,~N. Mag2Pol: a program for the analysis of spherical neutron polarimetry, flipping ratio and integrated intensity data. \emph{J. Appl. Cryst.} \textbf{2019}, \emph{52}, 175--185\relax
	\mciteBstWouldAddEndPuncttrue
	\mciteSetBstMidEndSepPunct{\mcitedefaultmidpunct}
	{\mcitedefaultendpunct}{\mcitedefaultseppunct}\relax
	\EndOfBibitem
	\bibitem[Blech and Averbach(1964)Blech, and Averbach]{Blech+Averbach1964}
	Blech,~I.~A.; Averbach,~B.~L. {Spin Correlations in MnO}. \emph{{P}hysics} \textbf{1964}, \emph{1}, 31--44\relax
	\mciteBstWouldAddEndPuncttrue
	\mciteSetBstMidEndSepPunct{\mcitedefaultmidpunct}
	{\mcitedefaultendpunct}{\mcitedefaultseppunct}\relax
	\EndOfBibitem
	\bibitem[Frandsen \latin{et~al.}(2022)Frandsen, Hamilton, Christensen, Stubben, and Billinge]{Frandsen+etal_mpdf_2022}
	Frandsen,~B.~A.; Hamilton,~P.~K.; Christensen,~J.~A.; Stubben,~E.; Billinge,~S. J.~L. diffpy.mpdf: open-source software for magnetic pair distribution function analysis. \emph{J. {A}ppl. {C}ryst.} \textbf{2022}, \emph{55}, 1377 -- 1382\relax
	\mciteBstWouldAddEndPuncttrue
	\mciteSetBstMidEndSepPunct{\mcitedefaultmidpunct}
	{\mcitedefaultendpunct}{\mcitedefaultseppunct}\relax
	\EndOfBibitem
	\bibitem[N\"agele \latin{et~al.}(1978)N\"agele, Knorr, Prandl, Convert, and Buevoz]{Naegele+etal1978}
	N\"agele,~W.; Knorr,~K.; Prandl,~W.; Convert,~P.; Buevoz,~J.~L. {Neutron scattering study of spin correlations and phase transitions in amorphous manganese aluminosilicates}. \emph{J. Phys. C: Solid State Phys.} \textbf{1978}, \emph{11}, 3295--3305\relax
	\mciteBstWouldAddEndPuncttrue
	\mciteSetBstMidEndSepPunct{\mcitedefaultmidpunct}
	{\mcitedefaultendpunct}{\mcitedefaultseppunct}\relax
	\EndOfBibitem
	\bibitem[Frandsen and Billinge(2015)Frandsen, and Billinge]{Frandsen+Billinge2015}
	Frandsen,~B.~A.; Billinge,~S. J.~L. Magnetic structure determination from the magnetic pair distribution function (mPDF): ground state of MnO. \emph{{A}cta {C}ryst.} \textbf{2015}, \emph{A71}, 325--334\relax
	\mciteBstWouldAddEndPuncttrue
	\mciteSetBstMidEndSepPunct{\mcitedefaultmidpunct}
	{\mcitedefaultendpunct}{\mcitedefaultseppunct}\relax
	\EndOfBibitem
	\bibitem[Roth \latin{et~al.}(2018)Roth, May, Ye, Chakoumakos, , and Iversen]{Roth+etal2018}
	Roth,~N.; May,~A.~F.; Ye,~F.; Chakoumakos,~B.~C.; ; Iversen,~B.~B. Model-free reconstruction of magnetic correlations in frustrated magnets. \emph{IUCrJ} \textbf{2018}, \emph{5}, 1 -- 7\relax
	\mciteBstWouldAddEndPuncttrue
	\mciteSetBstMidEndSepPunct{\mcitedefaultmidpunct}
	{\mcitedefaultendpunct}{\mcitedefaultseppunct}\relax
	\EndOfBibitem
	\bibitem[Patterson(1934)]{PattersonPR1934}
	Patterson,~A.~L. {A Fourier Series Method for the Determination of the Components of Interatomic Distances in Crystals}. \emph{Phys. Rev.} \textbf{1934}, \emph{46}, 372--376\relax
	\mciteBstWouldAddEndPuncttrue
	\mciteSetBstMidEndSepPunct{\mcitedefaultmidpunct}
	{\mcitedefaultendpunct}{\mcitedefaultseppunct}\relax
	\EndOfBibitem
	\bibitem[Patterson(1935)]{PattersonZfK1935}
	Patterson,~A.~L. {A direct method for the determination of the components of interatomic distances in crystals}. \emph{Zeitschrift f\"ur Kristallographie} \textbf{1935}, \emph{90}, 517\relax
	\mciteBstWouldAddEndPuncttrue
	\mciteSetBstMidEndSepPunct{\mcitedefaultmidpunct}
	{\mcitedefaultendpunct}{\mcitedefaultseppunct}\relax
	\EndOfBibitem
	\bibitem[Fischer \latin{et~al.}(2002)Fischer, Cuello, Palleau, Feltin, Barnes, Badyal, and Simonson]{Fischer+etal_D4c_2002}
	Fischer,~H.~E.; Cuello,~G.~J.; Palleau,~P.; Feltin,~D.; Barnes,~A.~C.; Badyal,~Y.~S.; Simonson,~J.~M. D4c: A very high precision diffractometer for disordered materials. \emph{Appl. Phys. A} \textbf{2002}, \emph{74}, s160 -- s162\relax
	\mciteBstWouldAddEndPuncttrue
	\mciteSetBstMidEndSepPunct{\mcitedefaultmidpunct}
	{\mcitedefaultendpunct}{\mcitedefaultseppunct}\relax
	\EndOfBibitem
	\bibitem[Fischer \latin{et~al.}(2000)Fischer, Palleau, and Feltin]{Fischer+etal_D4c_2000}
	Fischer,~H.~E.; Palleau,~P.; Feltin,~D. The D4c neutron diffractometer for liquids and glasses. \emph{Physica B} \textbf{2000}, \emph{93}, 276 -- 278\relax
	\mciteBstWouldAddEndPuncttrue
	\mciteSetBstMidEndSepPunct{\mcitedefaultmidpunct}
	{\mcitedefaultendpunct}{\mcitedefaultseppunct}\relax
	\EndOfBibitem
	\bibitem[Neuefeind \latin{et~al.}(2000)Neuefeind, Feygenson, Carruth, Hoffmann, and Chipley]{Neuefeind+etal_NOMAD_2012}
	Neuefeind,~J.; Feygenson,~M.; Carruth,~J.; Hoffmann,~R.; Chipley,~K.~K. The Nanoscale Ordered MAterials Diffractometer NOMAD at the Spallation Neutron Source SNS. \emph{Nuclear Instruments and Methods in Physics Research B} \textbf{2000}, \emph{287}, 68 -- 75\relax
	\mciteBstWouldAddEndPuncttrue
	\mciteSetBstMidEndSepPunct{\mcitedefaultmidpunct}
	{\mcitedefaultendpunct}{\mcitedefaultseppunct}\relax
	\EndOfBibitem
	\bibitem[Tsunoda \latin{et~al.}(2023)Tsunoda, Honda, Ikeda, Ohshita, Kambara, and Otomo]{Tsunoda+etal_NOVA_2023}
	Tsunoda,~M.; Honda,~T.; Ikeda,~K.; Ohshita,~H.; Kambara,~W.; Otomo,~T. Radial collimator performance and future collimator updates for the high-intensity total scattering diffractometer NOVA at J-PARC. \emph{Nucl. Instrum. Meth. A} \textbf{2023}, \emph{1055}, 168484\relax
	\mciteBstWouldAddEndPuncttrue
	\mciteSetBstMidEndSepPunct{\mcitedefaultmidpunct}
	{\mcitedefaultendpunct}{\mcitedefaultseppunct}\relax
	\EndOfBibitem
	\bibitem[Hannon(2005)]{Hannon+etal_GEM_2005}
	Hannon,~A.~C. Results on disordered materials from the GEneral Materials diffractometer. \emph{Nucl. Instrum. Meth. A} \textbf{2005}, \emph{551}, 88 -- 107\relax
	\mciteBstWouldAddEndPuncttrue
	\mciteSetBstMidEndSepPunct{\mcitedefaultmidpunct}
	{\mcitedefaultendpunct}{\mcitedefaultseppunct}\relax
	\EndOfBibitem
	\bibitem[Enderby \latin{et~al.}(2004)Enderby, Williams, Chapon, Hannon, Radaelli, and Soper]{Enderby+etal_GEM_2004}
	Enderby,~J.~E.; Williams,~W.~G.; Chapon,~L.~C.; Hannon,~A.~C.; Radaelli,~P.~G.; Soper,~A.~K. GEM: The general materials diffractometers at ISIS – multibank capabilities for studying crystalline and disordered materials. \emph{Neutron News} \textbf{2004}, \emph{15}, 19\relax
	\mciteBstWouldAddEndPuncttrue
	\mciteSetBstMidEndSepPunct{\mcitedefaultmidpunct}
	{\mcitedefaultendpunct}{\mcitedefaultseppunct}\relax
	\EndOfBibitem
	\bibitem[Bowron \latin{et~al.}(2010)Bowron, Soper, Jones, Ansell, Birch, Norris, Perrott, Riedel, Rhodes, Wakefield, Botti, Ricci, Grazzi, and Zoppi]{Bowron+etal_NIMROD_2010}
	Bowron,~D.~T.; Soper,~A.~K.; Jones,~K.; Ansell,~S.; Birch,~S.; Norris,~J.; Perrott,~L.; Riedel,~D.; Rhodes,~N.~J.; Wakefield,~S.~R.; Botti,~A.; Ricci,~M.-A.; Grazzi,~F.; Zoppi,~M. NIMROD: The Near and InterMediate Range Order Diffractometer of the ISIS second target station. \emph{Rev. Sci. Instrum.} \textbf{2010}, \emph{81}, 033905\relax
	\mciteBstWouldAddEndPuncttrue
	\mciteSetBstMidEndSepPunct{\mcitedefaultmidpunct}
	{\mcitedefaultendpunct}{\mcitedefaultseppunct}\relax
	\EndOfBibitem
	\bibitem[Huq \latin{et~al.}(2019)Huq, Kirkham, Peterson, Hodges, Whitfield, Page, H\"ugle, Iverson, Parizzi, and Rennich]{huq;jac19}
	Huq,~A.; Kirkham,~M.; Peterson,~P.~F.; Hodges,~J.~P.; Whitfield,~P.~S.; Page,~K.; H\"ugle,~T.; Iverson,~E.~B.; Parizzi,~A.; Rennich,~G. {POWGEN: rebuild of a third-generation powder diffractometer at the Spallation Neutron Source}. \emph{J. Appl. Cryst.} \textbf{2019}, \emph{52}, 1189--1201\relax
	\mciteBstWouldAddEndPuncttrue
	\mciteSetBstMidEndSepPunct{\mcitedefaultmidpunct}
	{\mcitedefaultendpunct}{\mcitedefaultseppunct}\relax
	\EndOfBibitem
	\bibitem[Hansen \latin{et~al.}(2008)Hansen, Henry, Fischer, Torregrossa, and Convert]{Hansen+etal_D4c_2008}
	Hansen,~T.~C.; Henry,~P.~F.; Fischer,~H.~E.; Torregrossa,~J.; Convert,~P. The D20 instrument at the ILL: a versatile high-intensity two-axis neutron diffractometer. \emph{Meas. Sci. Technol.} \textbf{2008}, \emph{19}, 034001\relax
	\mciteBstWouldAddEndPuncttrue
	\mciteSetBstMidEndSepPunct{\mcitedefaultmidpunct}
	{\mcitedefaultendpunct}{\mcitedefaultseppunct}\relax
	\EndOfBibitem
	\bibitem[Garlea \latin{et~al.}(2010)Garlea, Chakoumakos, Moore, Taylor, Chae, Maples, Riedel, Lynn, and Selby]{garle;apa10}
	Garlea,~V.~O.; Chakoumakos,~B.~C.; Moore,~S.~A.; Taylor,~G.~B.; Chae,~T.; Maples,~R.~G.; Riedel,~R.~A.; Lynn,~G.~W.; Selby,~D.~L. {The high-resolution powder diffractometer at the high flux isotope reactor}. \emph{Appl. Phys. A} \textbf{2010}, \emph{99}, 531--535\relax
	\mciteBstWouldAddEndPuncttrue
	\mciteSetBstMidEndSepPunct{\mcitedefaultmidpunct}
	{\mcitedefaultendpunct}{\mcitedefaultseppunct}\relax
	\EndOfBibitem
	\bibitem[Baral \latin{et~al.}(2024)Baral, Haglund, Liu, Kolesnikov, Mandrus, and Calder]{baral;prb24}
	Baral,~R.; Haglund,~A.~V.; Liu,~J.; Kolesnikov,~A.~I.; Mandrus,~D.; Calder,~S. {Local spin structure in the layered van der Waals materials MnPS$_x$Se$_{3-x}$}. \emph{Phys. Rev. B} \textbf{2024}, \emph{110}, 014423\relax
	\mciteBstWouldAddEndPuncttrue
	\mciteSetBstMidEndSepPunct{\mcitedefaultmidpunct}
	{\mcitedefaultendpunct}{\mcitedefaultseppunct}\relax
	\EndOfBibitem
	\bibitem[Zaliznyak \latin{et~al.}(2017)Zaliznyak, Savici, Garlea, Winn, Filges, Schneeloch, Tranquada, Gu, Wang, and Petrovic]{zaliz;jpconfs17}
	Zaliznyak,~I.~A.; Savici,~A.~T.; Garlea,~V.~O.; Winn,~B.; Filges,~U.; Schneeloch,~J.; Tranquada,~J.~M.; Gu,~G.; Wang,~A.; Petrovic,~C. Polarized neutron scattering on {HYSPEC}: the {HYbrid} {SPECtrometer} at {SNS}. \emph{J. Phys.: Conf. Ser.} \textbf{2017}, \emph{862}, 012030\relax
	\mciteBstWouldAddEndPuncttrue
	\mciteSetBstMidEndSepPunct{\mcitedefaultmidpunct}
	{\mcitedefaultendpunct}{\mcitedefaultseppunct}\relax
	\EndOfBibitem
	\bibitem[Frandsen \latin{et~al.}(2022)Frandsen, Baral, Winn, and Garlea]{frand;jap22}
	Frandsen,~B.~A.; Baral,~R.; Winn,~B.; Garlea,~V.~O. {Magnetic pair distribution function data using polarized neutrons and \textit{ad hoc} corrections}. \emph{J. Appl. Phys.} \textbf{2022}, \emph{132}, 223909\relax
	\mciteBstWouldAddEndPuncttrue
	\mciteSetBstMidEndSepPunct{\mcitedefaultmidpunct}
	{\mcitedefaultendpunct}{\mcitedefaultseppunct}\relax
	\EndOfBibitem
	\bibitem[Juh\'{a}s \latin{et~al.}(2015)Juh\'{a}s, Farrow, Yang, Knox, and Billinge]{juhas;aca15}
	Juh\'{a}s,~P.; Farrow,~C.~L.; Yang,~X.; Knox,~K.~R.; Billinge,~S. J.~L. {Complex Modeling: a strategy and software program for combining multiple information sources to solve ill-posed structure and nanostructure inverse problems}. \emph{{A}cta {C}ryst.} \textbf{2015}, \emph{A71}, 562--568\relax
	\mciteBstWouldAddEndPuncttrue
	\mciteSetBstMidEndSepPunct{\mcitedefaultmidpunct}
	{\mcitedefaultendpunct}{\mcitedefaultseppunct}\relax
	\EndOfBibitem
	\bibitem[Tucker \latin{et~al.}(2007)Tucker, Keen, Dove, Goodwin, and Hui]{Tucker+etal_RMCProfile_2007}
	Tucker,~M.~G.; Keen,~D.~A.; Dove,~M.~T.; Goodwin,~A.~L.; Hui,~Q. RMCProfile: reverse Monte Carlo for polycrystalline materials. \emph{J. Phys.: Condens. Matter} \textbf{2007}, \emph{19}, 335218\relax
	\mciteBstWouldAddEndPuncttrue
	\mciteSetBstMidEndSepPunct{\mcitedefaultmidpunct}
	{\mcitedefaultendpunct}{\mcitedefaultseppunct}\relax
	\EndOfBibitem
	\bibitem[Ren \latin{et~al.}(2017)Ren, Yang, Jiang, Zhang, Zhou, Li, Xin, and He]{ren;jmchemc17}
	Ren,~Y.; Yang,~J.; Jiang,~Q.; Zhang,~D.; Zhou,~Z.; Li,~X.; Xin,~J.; He,~X. {Synergistic effect by Na doping and S substitution for high thermoelectric performance of p-type MnTe}. \emph{J. Mater. chem. C} \textbf{2017}, \emph{5}, 5076--5082\relax
	\mciteBstWouldAddEndPuncttrue
	\mciteSetBstMidEndSepPunct{\mcitedefaultmidpunct}
	{\mcitedefaultendpunct}{\mcitedefaultseppunct}\relax
	\EndOfBibitem
	\bibitem[Xu \latin{et~al.}(2017)Xu, Li, Wang, Li, Chen, Lin, Chen, and Pei]{xu;jmchema17}
	Xu,~Y.; Li,~W.; Wang,~C.; Li,~J.; Chen,~Z.; Lin,~S.; Chen,~Y.; Pei,~Y. {Performance optimization and single parabolic band behavior of thermoelectric MnTe}. \emph{J. Mater. Chem. A} \textbf{2017}, \emph{5}, 19143--19150\relax
	\mciteBstWouldAddEndPuncttrue
	\mciteSetBstMidEndSepPunct{\mcitedefaultmidpunct}
	{\mcitedefaultendpunct}{\mcitedefaultseppunct}\relax
	\EndOfBibitem
	\bibitem[Zulkifal \latin{et~al.}(2023)Zulkifal, Wang, Zhang, Siddique, Yu, Wang, Gong, Li, Li, Zhang, Wang, and Tang]{zulki;advs23}
	Zulkifal,~S.; Wang,~Z.; Zhang,~X.; Siddique,~S.; Yu,~Y.; Wang,~C.; Gong,~Y.; Li,~S.; Li,~D.; Zhang,~Y.; Wang,~P.; Tang,~G. {Multiple Valence Bands Convergence and Localized Lattice Engineering Lead to Superhigh Thermoelectric Figure of Merit in MnTe}. \emph{Adv. Sci.} \textbf{2023}, 2206342\relax
	\mciteBstWouldAddEndPuncttrue
	\mciteSetBstMidEndSepPunct{\mcitedefaultmidpunct}
	{\mcitedefaultendpunct}{\mcitedefaultseppunct}\relax
	\EndOfBibitem
	\bibitem[Kriegner \latin{et~al.}(2016)Kriegner, V\'yborn\'y, Olejn\'ik, Reichlov\'a, Nov\'ak, Marti, Gazquez, Saidl, N\v{e}mec, Volobuev, Springholz, Hol\'y, and Jungwirth]{krieg;nc16}
	Kriegner,~D.; V\'yborn\'y,~K.; Olejn\'ik,~K.; Reichlov\'a,~H.; Nov\'ak,~V.; Marti,~X.; Gazquez,~J.; Saidl,~V.; N\v{e}mec,~P.; Volobuev,~V.~V.; Springholz,~G.; Hol\'y,~V.; Jungwirth,~T. {Multiple-stable anisotropic magnetoresistance memory in antiferromagnetic MnTe}. \emph{Nat. Commun.} \textbf{2016}, \emph{7}, 11623\relax
	\mciteBstWouldAddEndPuncttrue
	\mciteSetBstMidEndSepPunct{\mcitedefaultmidpunct}
	{\mcitedefaultendpunct}{\mcitedefaultseppunct}\relax
	\EndOfBibitem
	\bibitem[Mazin(2023)]{mazin;prb23}
	Mazin,~I.~I. {Altermagnetism in MnTe: Origin, predicted manifestations, and routes to detwinning}. \emph{Phys. Rev. B} \textbf{2023}, \emph{107}, L100418\relax
	\mciteBstWouldAddEndPuncttrue
	\mciteSetBstMidEndSepPunct{\mcitedefaultmidpunct}
	{\mcitedefaultendpunct}{\mcitedefaultseppunct}\relax
	\EndOfBibitem
	\bibitem[Krempask\'y \latin{et~al.}(2024)Krempask\'y, \v{S}mejkal, D'Souza, Hajlaoui, Springholz, Uhl\'i\v{r}ov\'a, Alarab, Constantinou, Strocov, Usanov, Pudelko, Gonz\'alez-Hern\'andez, Birk~Hellenes, Jansa, Reichlov\'a, \v{S}ob\'a\v{n}, Gonzalez~Betancourt, Wadley, Sinova, Kriegner, Min\'ar, Dil, and Jungwirth]{kremp;n24}
	Krempask\'y,~J. \latin{et~al.}  {Altermagnetic lifting of Kramers spin degeneracy}. \emph{Nature} \textbf{2024}, \emph{626}, 517--522\relax
	\mciteBstWouldAddEndPuncttrue
	\mciteSetBstMidEndSepPunct{\mcitedefaultmidpunct}
	{\mcitedefaultendpunct}{\mcitedefaultseppunct}\relax
	\EndOfBibitem
	\bibitem[Baral \latin{et~al.}(2023)Baral, Abeykoon, Campbell, and Frandsen]{baral;afm23}
	Baral,~R.; Abeykoon,~A.~M.; Campbell,~B.~J.; Frandsen,~B.~A. {Giant Spontaneous Magnetostriction in MnTe Driven by a Novel Magnetostructural Coupling Mechanism}. \emph{Adv. Funct. Mater.} \textbf{2023}, \emph{33}, 2305247\relax
	\mciteBstWouldAddEndPuncttrue
	\mciteSetBstMidEndSepPunct{\mcitedefaultmidpunct}
	{\mcitedefaultendpunct}{\mcitedefaultseppunct}\relax
	\EndOfBibitem
	\bibitem[Lefran\c{c}ois \latin{et~al.}(2019)Lefran\c{c}ois, Mangin-Thro, Lhotel, Robert, Petit, Cathelin, Fischer, Colin, Damay, Ollivier, Lejay, Chapon, Simonet, and Ballou]{Lefrancois+etal2019}
	Lefran\c{c}ois,~E.; Mangin-Thro,~L.; Lhotel,~E.; Robert,~J.; Petit,~S.; Cathelin,~V.; Fischer,~H.~E.; Colin,~C.~V.; Damay,~F.; Ollivier,~J.; Lejay,~P.; Chapon,~L.~C.; Simonet,~V.; Ballou,~R. Spin decoupling under a staggered field in the \ce{Gd2Ir2O7} pyrochlore. \emph{{P}hys. {R}ev. {B}} \textbf{2019}, \emph{99}, 060401(R)\relax
	\mciteBstWouldAddEndPuncttrue
	\mciteSetBstMidEndSepPunct{\mcitedefaultmidpunct}
	{\mcitedefaultendpunct}{\mcitedefaultseppunct}\relax
	\EndOfBibitem
	\bibitem[Roth \latin{et~al.}(2019)Roth, Ye, May, Chakoumakos, and Iversen]{Roth+etal2019}
	Roth,~N.; Ye,~F.; May,~A.~F.; Chakoumakos,~B.~C.; Iversen,~B.~B. Magnetic correlations and structure in bixbyite across the spin-glass transition. \emph{{P}hys. {R}ev. {B}} \textbf{2019}, \emph{100}, 144404\relax
	\mciteBstWouldAddEndPuncttrue
	\mciteSetBstMidEndSepPunct{\mcitedefaultmidpunct}
	{\mcitedefaultendpunct}{\mcitedefaultseppunct}\relax
	\EndOfBibitem
	\bibitem[Dun \latin{et~al.}(2021)Dun, Daum, Baral, Fischer, Cao, Liu, Stone, Rodriguez-Rivera, Choi, Huang, Zhou, Mourigal, and Frandsen]{Dun+etal2021}
	Dun,~Z.; Daum,~M.; Baral,~R.; Fischer,~H.~E.; Cao,~H.; Liu,~Y.; Stone,~M.~B.; Rodriguez-Rivera,~J.~A.; Choi,~E.~S.; Huang,~Q.; Zhou,~H.; Mourigal,~M.; Frandsen,~B.~A. Neutron scattering investigation of proposed Kosterlitz-Thouless transitions in the triangular-lattice Ising antiferromagnet \ce{TmMgGaO4}. \emph{{P}hys. {R}ev. {B}} \textbf{2021}, \emph{103}, 064424\relax
	\mciteBstWouldAddEndPuncttrue
	\mciteSetBstMidEndSepPunct{\mcitedefaultmidpunct}
	{\mcitedefaultendpunct}{\mcitedefaultseppunct}\relax
	\EndOfBibitem
	\bibitem[Singamaneni \latin{et~al.}(2011)Singamaneni, Bliznyuk, Binek, and Tsymbal]{singa;jmc11}
	Singamaneni,~S.; Bliznyuk,~V.~N.; Binek,~C.; Tsymbal,~E.~Y. {Magnetic nanoparticles: recent advances in synthesis, self-assembly and applications}. \emph{J. Mater. Chem.} \textbf{2011}, \emph{21}, 16819--16845\relax
	\mciteBstWouldAddEndPuncttrue
	\mciteSetBstMidEndSepPunct{\mcitedefaultmidpunct}
	{\mcitedefaultendpunct}{\mcitedefaultseppunct}\relax
	\EndOfBibitem
	\bibitem[Majetich \latin{et~al.}(2011)Majetich, Wen, and Booth]{majet;acsn11}
	Majetich,~S.~A.; Wen,~T.; Booth,~R.~A. {Functional Magnetic Nanoparticle Assemblies: Formation, Collective Behavior, and Future Directions}. \emph{ACS Nano} \textbf{2011}, \emph{5}, 6081--6084\relax
	\mciteBstWouldAddEndPuncttrue
	\mciteSetBstMidEndSepPunct{\mcitedefaultmidpunct}
	{\mcitedefaultendpunct}{\mcitedefaultseppunct}\relax
	\EndOfBibitem
	\bibitem[Mater\'on \latin{et~al.}(2021)Mater\'on, Miyazaki, Carr, Joshi, Picciani, Dalmaschio, Davis, and Shimizu]{mater;assa21}
	Mater\'on,~E.~M.; Miyazaki,~C.~M.; Carr,~O.; Joshi,~N.; Picciani,~P.~H.; Dalmaschio,~C.~J.; Davis,~F.; Shimizu,~F.~M. {Magnetic nanoparticles in biomedical applications: A review}. \emph{Appl. Surf. Sci. Adv.} \textbf{2021}, \emph{6}, 100163\relax
	\mciteBstWouldAddEndPuncttrue
	\mciteSetBstMidEndSepPunct{\mcitedefaultmidpunct}
	{\mcitedefaultendpunct}{\mcitedefaultseppunct}\relax
	\EndOfBibitem
	\bibitem[kianfar(2021)]{kianf;jsnm21}
	kianfar,~E. {Magnetic Nanoparticles in Targeted Drug Delivery: a Review}. \emph{J. Supercond. Novel Magn.} \textbf{2021}, \emph{34}, 1709--1735\relax
	\mciteBstWouldAddEndPuncttrue
	\mciteSetBstMidEndSepPunct{\mcitedefaultmidpunct}
	{\mcitedefaultendpunct}{\mcitedefaultseppunct}\relax
	\EndOfBibitem
	\bibitem[Fatima \latin{et~al.}(2021)Fatima, Charinpanitkul, and Kim]{fatim;nanom21}
	Fatima,~H.; Charinpanitkul,~T.; Kim,~K.-S. {Fundamentals to Apply Magnetic Nanoparticles for Hyperthermia Therapy}. \emph{Nanomaterials} \textbf{2021}, \emph{11}, 1203\relax
	\mciteBstWouldAddEndPuncttrue
	\mciteSetBstMidEndSepPunct{\mcitedefaultmidpunct}
	{\mcitedefaultendpunct}{\mcitedefaultseppunct}\relax
	\EndOfBibitem
	\bibitem[Na \latin{et~al.}(2009)Na, Song, and Hyeon]{na;am09}
	Na,~H.~B.; Song,~I.~C.; Hyeon,~T. {Inorganic Nanoparticles for MRI Contrast Agents}. \emph{Adv. Mater.} \textbf{2009}, \emph{21}, 2133--2148\relax
	\mciteBstWouldAddEndPuncttrue
	\mciteSetBstMidEndSepPunct{\mcitedefaultmidpunct}
	{\mcitedefaultendpunct}{\mcitedefaultseppunct}\relax
	\EndOfBibitem
	\bibitem[Andersen \latin{et~al.}(2021)Andersen, Frandsen, Gunnlaugsson, J{\o}rgensen, Billinge, Jensen, and Christensen]{ander;ij21}
	Andersen,~H.~L.; Frandsen,~B.~A.; Gunnlaugsson,~H.~P.; J{\o}rgensen,~M. R.~V.; Billinge,~S. J.~L.; Jensen,~K. M.~{\O}.; Christensen,~M. {Local and long-range atomic/magnetic structure of non-stoichiometric spinel iron oxide nanocrystallites}. \emph{IUCrJ} \textbf{2021}, \emph{8}, 33--45\relax
	\mciteBstWouldAddEndPuncttrue
	\mciteSetBstMidEndSepPunct{\mcitedefaultmidpunct}
	{\mcitedefaultendpunct}{\mcitedefaultseppunct}\relax
	\EndOfBibitem
	\bibitem[Frison \latin{et~al.}(2013)Frison, Cernuto, Cervellino, Zaharko, Colonna, Guagliardi, and Masciocchi]{friso;cm13}
	Frison,~R.; Cernuto,~G.; Cervellino,~A.; Zaharko,~O.; Colonna,~G.~M.; Guagliardi,~A.; Masciocchi,~N. {Magnetite–Maghemite Nanoparticles in the 5–15 nm Range: Correlating the Core–Shell Composition and the Surface Structure to the Magnetic Properties. A Total Scattering Study.} \emph{Chem. Mater.} \textbf{2013}, \emph{25}, 4820--4827\relax
	\mciteBstWouldAddEndPuncttrue
	\mciteSetBstMidEndSepPunct{\mcitedefaultmidpunct}
	{\mcitedefaultendpunct}{\mcitedefaultseppunct}\relax
	\EndOfBibitem
	\bibitem[Iida \latin{et~al.}(2022)Iida, Kodama, Inamura, Nakamura, Chang, and Shamoto]{Iida+etal2022}
	Iida,~K.; Kodama,~K.; Inamura,~Y.; Nakamura,~M.; Chang,~L.; Shamoto,~S.-i. Magnon mode transition in real space. \emph{Nature Scientific Reports} \textbf{2022}, \emph{12}, 20663\relax
	\mciteBstWouldAddEndPuncttrue
	\mciteSetBstMidEndSepPunct{\mcitedefaultmidpunct}
	{\mcitedefaultendpunct}{\mcitedefaultseppunct}\relax
	\EndOfBibitem
\end{mcitethebibliography}
\providecommand{\latin}[1]{#1}
\makeatletter
\providecommand{\doi}
{\begingroup\let\do\@makeother\dospecials
	\catcode`\{=1 \catcode`\}=2 \doi@aux}
\providecommand{\doi@aux}[1]{\endgroup\texttt{#1}}
\makeatother
\providecommand*\mcitethebibliography{\thebibliography}
\csname @ifundefined\endcsname{endmcitethebibliography}  {\let\endmcitethebibliography\endthebibliography}{}

\end{document}